\documentclass[11pt]{article}

\usepackage{jheppub} 

\usepackage{braket,slashed,bm}
\usepackage{array,multirow}
\usepackage[normalem]{ulem}
\usepackage[T1]{fontenc} 

\usepackage{mathrsfs}

\usepackage{braket,slashed,bm}
\usepackage{array,multirow}
\usepackage[normalem]{ulem}
\usepackage[T1]{fontenc}
\usepackage{mathrsfs}
\usepackage{tikz}
\usetikzlibrary{arrows,decorations.pathmorphing,backgrounds,positioning,fit,petri,automata,shadows,calendar,mindmap,
decorations.markings,calc}

\def\nn{\nonumber\\ }

\def\lsix{ \mathcal{L}^{(6)}}

\def\hyp{\mathsf{y}}

\def\lsix{ \mathcal{L}^{(6)}}
\def\gcb{{\overline g_{1}}}
\def\gcw{{\overline g_{2}}}

\def\tc{{\overline \theta}}
\def\that{{\hat \theta}}

\def\sc{{\overline s}}

\def\bea{\begin{eqnarray}}
\def\eea{\end{eqnarray}}

\title{
Towards consistent Electroweak Precision Data constraints in the SMEFT}

\author{
Laure Berthier and Michael Trott\\
Niels Bohr International Academy,
University of Copenhagen, Blegdamsvej 17, DK-2100 Copenhagen, Denmark \\
bethier@nbi.ku.dk, michael.trott@cern.ch}

\abstract{We discuss the impact of many previously neglected effects of higher
dimensional operators when fitting to Electroweak Precision data (EWPD) in the Standard Model Effective Field Theory (SMEFT).
We calculate the general case of $2 \rightarrow 2$ fermion scattering in the SMEFT to order $\mathcal{O}(\bar{v}_T^2/\Lambda^2)$
valid on and off the $Z$ pole, in the massless fermion limit. We demonstrate that previously neglected corrections scale as $\Gamma_Z M_Z/\bar{v}_T^2$ in the partial widths
extracted from measured cross sections at LEPI, compared to the leading effect of dimension six operators in anomalous $Z$ couplings.
Further, constraints on leading effects of anomalous $Z$ couplings are also modified
by neglected perturbative corrections and dimension eight operators.  We perform a minimal EWPD fit
to illustrate the size of the error these corrections induce, when bounding leading effects.  These considerations relax bounds
compared to a naive leading order analysis, and show that constraints that rise above the percent level are subject to substantial theoretical uncertanties.
We also argue that renormalization group running global constraints expressed through $\chi^2$ functions to a common scale,
and then minimizing and performing a global fit of all data allows more consistent constraints to be obtained in the SMEFT.}
\begin{document}
\maketitle
\section{Introduction}\label{sec:intro}

The discovery of a Higgs like scalar at LHC, with couplings to the $\rm W^\pm,Z$
in rough agreement with the Standard Model (SM) expectation, allows the cut off scale of the Standard Model Effective Field Theory (SMEFT)
to be further separated from the electroweak vaccum expectation value ($\bar{v}_T$) in the SMEFT.\footnote{Compared to the case where no $0^+$ scalar is present.} Nevertheless, expectations of naturalness still motivate precision
studies of the SMEFT. The aim is to search for patterns of deviations that could be present as the low energy footprint of beyond the Standard Model (BSM) physics.
Further, a precise knowledge of the global constraint picture of the SMEFT is crucially important to place
any discovered state at LHC into the proper experimental context, including the discovered $0^+$ scalar.  The purpose of this
paper is to advance this effort, by further developing the analysis of model independent global constraints on the SMEFT.

Determining the global constraint picture in the general linear SMEFT is a challenge, due to the complicated nature of this theory.\footnote{For some past global constraint analyses and comments relevant to this work see Ref \cite{Grinstein:1991cd,Han:2004az,Ciuchini:2013pca,Chen:2013kfa,Ciuchini:2014dea,Baak:2014ora,Chen:2013kfa,Durieux:2014xla,Petrov:2015jea,Wells:2014pga,Ellis:2014dva,Trott:2014dma,Henning:2014wua,Pomarol:2013zra,Falkowski:2014tna}.
The complexity of the theory is perhaps best illustrated by the fact that
the non redundant basis of dimension-six operators in the (linear) SMEFT given in Ref.~\cite{Grzadkowski:2010es}
has 2499 parameters \cite{Alonso:2013hga}.} The linear SMEFT is defined by the assumption that the low energy limit of
BSM physics is adequately described by an EFT that assumes the observed $0^+$ scalar is embedded in the Higgs doublet, with the addition of higher dimensional operators ($\mathcal{L}^{(5)}+ \lsix + \cdots$) constructed out of the $\rm SU(3) \times SU(2) \times U(1)$ invariant SM fields. This is the assumption we adopt in this
paper. Based on this choice, $\lsix$ has been classified in Refs.~\cite{Buchmuller:1985jz,Grzadkowski:2010es}.
Recently,  $\mathcal{L}^{(7)}$ has  been classified in Ref \cite{Lehman:2014jma}. We will restrict our attention to the dimension
six lepton and baryon number conserving operator corrections to the linear SMEFT in this paper, except when dimension eight operators are used to
characterize theoretical errors. Note that the dimension seven operators violate Lepton number \cite{Lehman:2014jma}, as does $\mathcal{L}^{(5)}$, and as such, these operators can be neglected for our purposes, and not included in theoretical error estimates.

In this paper we advance the understanding of the global constraints on the linear SMEFT due to near $Z$ pole data.\footnote{The qualifier "near"
the $Z$ pole is important as some interference effects vanish when data is taken exactly on the $Z$ pole.
At LEPI a significant fraction of data (approximately 1/4th) is taken off the $Z$ pole
to fit for the $Z$ mass, total width and cross section as a function of center of mass collision energy $s$. The combined
data set includes this off pole data (approximately corrected to account for off pole $\gamma-Z$ interference effects in the SM).
See Ref. \cite{Z-Pole} for a description of the LEPI program.
LEPII was run far off the $Z$ pole.}
We calculate $d \sigma(\ell^+ \ell^- \rightarrow \bar{f} f)/ d \cos \theta $
where $f = \{e,\mu,\tau,u, c,b,s,d\} $ on and off the $Z$ pole in the massless limit,  to order $\mathcal{O}(\bar{v}_T^2/\Lambda^2)$
in the SMEFT.
We emphasize the need for consistency in how these processes are treated, and point out several corrections of $\lsix$ to the SM that have been neglected in past global constraint efforts.

Our main point is the following. When considering constraints on $\lsix$, theoretical calculations are never performed to arbitrary precision.
As a result, bounds on $\lsix$ in a purely leading order analysis (of BSM effects) can not rise to an arbitrary level of
constraint in a self consistent way. Terms that are sub-leading in the power counting of the EFT are neglected. Loop corrections involving higher dimensional operators are
also generally neglected when considering Electroweak precision data (EWPD).
Further, the contributions of BSM effects in processes that are sub-leading {\it in the SM} have also been neglected.

All of these assumptions are potentially problematic
for consistent analyses, when very strong bounds are argued to be obtained on $\lsix$ in a naive leading order analysis.
In this paper, we argue that EWPD bounds on anomalous $Z$ couplings that exceed the percent level are challenged due to this litany of
neglected corrections. The up side of considering sub-leading corrections more consistently in the SMEFT
is a relaxing of bounds on $\lsix$, when a truly general analysis is performed.

We discuss these issues and a more consistent approach to EWPD on and off the  $Z$ pole in the SMEFT in Section \ref{obs}.
It is essential to eventually also include the less precise
results of off $Z$ pole data reported in LEPII in a global analysis of the SMEFT. Our results are general enough to perform this analysis for LEPII data.
In such an effort, some of the interference effects that we highlight are only suppressed compared to the leading order terms by $M_Z^2/\bar{v}_T^2$. For near $Z$ pole data
these interference effects scale as $\Gamma_Z^2/M_Z^2$ (for $\gamma-Z$ corrections) and $\Gamma_Z^2/\bar{v}_T^2$ (for $\psi^4-Z$ corrections) in the $2 \rightarrow 2$ scattering cross sections. However, the latter effects lead to corrections relatively suppressed by $\Gamma_Z \, M_Z/\bar{v}_T^2$, compared to the leading effects of dimension six operators, in the partial widths inferred from these cross sections. These corrections vanish when the cross sections are measured exactly on the $Z$ pole,
which holds for the majority, but not the totality, of the global LEP1 data set.

The majority of our results are general enough that we need not impose a $\rm U(3)^5$ flavour symmetry assumption on
the dimension six operators in the SMEFT. In some particular cases, we will make the
simplifying assumption that any beyond the SM flavour violation follows a linear minimal flavour violation (MFV)
hypothesis \cite{Chivukula:1987py,Hall:1990ac,DAmbrosio:2002ex,Buras:2000dm}
consistent with $\rm U(3)^5$ flavour symmetry.
In this case,  the flavour structure of the dimension six operators of the SMEFT is trivialized down to the case where
only $76$ parameters are present \cite{Alonso:2013hga}.

The outline of this paper is as  follows. In Section \ref{PC} we discuss the power counting we employ.
In Sections \ref{sec:input},\ref{vectorcouplings} we review the reformulation of the input parameters used in predictions in the SMEFT.
In Section \ref{differential} we report the differential cross sections for $2 \rightarrow 2$ scattering consistently
generalized into the SMEFT to leading order in dimension six operators. In Section \ref{partialwidths} we discuss how the
near pole cross sections used to infer partial widths, when generalized consistently in the SMEFT, receive corrections
that are relevant to $\mathcal{O}(10^{-3})$ bounds on $\lsix$ effects that modify $Z \bar{f} \, f$.
In Section \ref{Numerics} we illustrate the impact of these previously neglected corrections on extractions of the bounds on $\lsix$.
 We then argue that renormalization group (RG) running  a global EWPD constraint function directly to the energy scales relevant for LHC processes
is preferred, in order to obtain accurate constraints in the linear SMEFT. In Section \ref{conclusions} we conclude.

\subsection{Power counting}\label{PC}
The relative importance of various local operators in the SMEFT depends on the power counting, and the
particular Wilson coefficient that an operator obtains when matching onto an unknown BSM sector.\footnote{Conflating these two issues
by suppressing operators by $1/\bar{v}_T^2$ and absorbing all suppression into a modified Wilson coefficient is a challenge for any consistent power
counting scheme. Such an approach can lead to the EFT being used beyond its regime of validity - set by the suppression scale present in the power counting, $\Lambda$.}
In the SMEFT, the most naive and general power counting is to assign each dimension six operator
a suppression by $1/\Lambda^2$ and to retain all operators up to a fixed order in $1/\Lambda$.

Alternate approaches to utilizing this naive power counting exist in the literature. A prominent example is the  Naive Dimensional Analysis (NDA) approach laid out in Ref \cite{Manohar:1983md}. NDA was
developed by examining the consistency of the chiral quark model, but has been found to be broadly applicable in other applications.
NDA  states that an operator generated
at the scale $\Lambda$ in an EFT can be written as
\begin{align}
f^2 \Lambda^2 \left( \frac{H}{f} \right)^A \left( \frac{\psi}{f \sqrt \Lambda} \right)^B   \left( \frac{g X}{\Lambda^2} \right)^C
 \left( \frac{D}{\Lambda} \right)^D\,,
 \label{nda}
\end{align}
with the {\it approximate} identification $\Lambda \sim 4\pi f$. Here $H$ is a scalar field, $\psi$ is a general chiral fermion field, $X$ is a general gauge field strength tensor
with corresponding gauge coupling $g$. The powers $A,B,C,D$ correspond to the number of the corresponding fields present in a particular operator.
Recently it has been shown that the NDA scheme is incomplete in some scenarios, but it can be consistently extended \cite{Jenkins:2013sda,Buchalla:2013eza}.
In what follows, we emphasize the need for the consistent inclusion of four fermion ($\psi^4$) operators in EWPD, and the effect of including these operators when bounds on terms in $\lsix$ of the form $H D H \, \psi^2$
are obtained. We note that both these operator classes have the same scaling in NDA.

Other schemes have also been proposed. For some weakly coupled renormalizable UV models generating higher dimensional operators, an analysis based on when operators
can be obtained in a matching at tree or loop level was developed in Ref \cite{Arzt:1994gp}, and can be self-consistent. Yet another approach distinct from this classification
is discussed in Ref \cite{Heinemeyer:2013tqa}. For some discussion on the claims of this latter scheme, see  Ref \cite{Buchalla:2014eca,Jenkins:2013fya}.

A truly general power counting scheme that is valid for all possible UV models, covering the cases of both weakly and strongly interacting, and allowing the UV to be an EFT itself, would be suitable to utilize in the SMEFT. Due to the absence of such a scheme, we naively suppress all dimension six operators by $1/\Lambda^2$. With this power counting, the case $\Lambda \sim \rm TeV$ is of most interest, so that  $\bar{v}_T^2/\Lambda^2 \sim 10^{-2}$.
Naively incorporating a per-mille constraint in EWPD on a combination of dimension six Wilson coefficients, denoted $c_6$, corresponds to $c_6 \, \bar{v}_T^2/\Lambda^2 \lesssim 10^{-3}$, which gives $c_6 \lesssim 0.1$ for $\Lambda \sim  \, 2.5 \, {\rm TeV}$. Such a bound generally neglects the effects of the large number of un-numerated (and even undefined) dimension eight operators in the SMEFT. So that schematically $c_6 + 0.01 \, c_8 \lesssim 0.1$ for $\rm TeV$ cut off scales. Bounds of this form are difficult to consider as precise numerical limits on the inferred Wilson coefficients. We will return to this point in Section \ref{Numerics}.

\section{Electroweak Parameters}\label{sec:input}
The approach we take in this paper is to more consistently generalize the predictions in the SM to the
SMEFT.\footnote{For the case of a minimal oblique parameter analysis of EWPD, the basic ideas of the approach we employ are reviewed
in Ref \cite{Wells:2005vk}.} To construct theoretical predictions of EWPD,
we take as core input parameters for the Electroweak sector the measured values
of the fine structure constant $\hat{\alpha}_{ew}$ from the low energy limit of electron Compton scattering, the Fermi decay constant in muon decays $\hat{G}_F$ and the measured $Z$ mass ($\hat{m}_Z$). It is convenient to relate observables in terms of the parameters $g_2, \sin^2 \theta = g_1^2/(g_1^2 + g_2^2)$ and the electroweak vacuum expectation value (vev) {\it{v}}.
Defining at tree level the effective {\it measured} mixing angle
\bea\label{sinequation}
\sin^2 \hat{\theta} = \frac{1}{2} - \frac{1}{2}\sqrt{1 - \frac{4 \, \pi \hat{\alpha}_{ew}}{\sqrt{2} \, \hat{G}_F \, \hat{m}_Z^2}},
\eea
then the measured value of the $\rm SU_L(2)$ gauge coupling can be inferred (at tree level) via
\bea
 \hat{g}_2 \, \sin \hat{\theta} = 2 \, \sqrt{\pi} \, \hat{\alpha}_{ew}^{1/2}.
\eea
The effective measured vacuum expectation value (vev) in the SM can be defined as $\hat{v}^2 = 1/\sqrt{2} \, \hat{G}_F$.
 All of these input parameters
are redefined going from the SM to the SMEFT, and the resulting shifts are characterized in Section \ref{inputs}.
We will consistently use the notation that the measured parameters, or inferred measured parameters (such as $\sin^2 \hat{\theta},\hat{g}_2$), are denoted with
a hat superscript. In relating predictions to these input parameters we will consistently only include corrections in the SMEFT
that are suppressed by $\bar{v}_T^2/\Lambda^2$, neglecting  $\bar{v}_T^4/\Lambda^4$ contributions. For this reason SMEFT parameters multiplying insertions of higher dimensional operators
can be traded for $\hat{\alpha}_{ew},\hat{v}^2,\hat{m}_Z$ using the SM relations.\footnote{As well as these core input parameters, we also note that the values of $\{m_t, \alpha_s, m_H, m_c, m_b, m_\tau,V_{CKM}^{ij}, \Delta \alpha_{had}^{(5)}, \cdots \}$ are also
required in a truly global EWPD analysis of all data.}
\begin{center}
\begin{table}
\centering
\tabcolsep 8pt
\begin{tabular}{|c|c|c|}
\hline
Parameter & Input Value & Ref.  \\ \hline
$\hat{m}_Z$ & $91.1875 \pm 0.0021$ & \cite{Z-Pole,Agashe:2014kda,Mohr:2012tt} \\
$\hat{G}_F$ & $1.1663787(6) \times 10^{-5} $ &  \cite{Agashe:2014kda,Mohr:2012tt} \\
$\hat{\alpha}_{ew}$ & $1/137.035999074(94) $ &  \cite{Agashe:2014kda,Mohr:2012tt} \\
\hline
\end{tabular}
\caption{Current best estimates of the core input parameters used to make predictions in the SMEFT. }
\end{table}
\end{center}
\subsection{Input Parameters}\label{inputs}

Calculating expressions, we use the canonically normalized SMEFT in the basis of Ref. \cite{Grzadkowski:2010es}.
By canonically normalized, we mean that the kinetic terms of all propagating fields have been taken to a minimal
form, with a field and $\bar{v}_T^2$ independent Wilson coefficient.
Many of our results build upon the discussion in Ref.\cite{Alonso:2013hga}. For example, the canonically normalized SMEFT
Lagrangian parameters are denoted with bar superscripts, as defined in Ref.\cite{Alonso:2013hga}. The SM Lagrangian parameters and theoretical predictions for observables in the SM will have no superscript (no hat and no bar) and if we stop at the leading order of the SM value we will add : $(...)_{SM}$ to specify it.
In the following Sections we will use the shorthand notation $s_{\hat{\theta}}^2 = \sin^2 \hat{\theta}$,
$c_{\hat{\theta}}^2 = \cos^2 \hat{\theta}$.\footnote{See the Appendix for a discussion of the notational conventions.} The canonically normalized gauge fields introduce the gauge couplings given by $g_{1,2} = \bar{g}_{1,2}  (1 + C_{H(B,W)} \, \bar{v}_T^2)$.
For completeness, we summarize the relation between
the SMEFT  Lagrangian parameters and the  measured input parameters in this Section.

\subsubsection{$G_F$}
We define the local effective interaction for muon decay as
\begin{align}
\mathcal{L}_{G_F} =  -\frac{4\hat{\mathcal{G}}_F}{\sqrt{2}} \, \left(\bar{\nu}_\mu \, \gamma^\mu P_L \mu \right) \left(\bar{e} \, \gamma_\mu P_L \nu_e\right).
\end{align}
The parameter $\hat{\mathcal{G}}_F$ is fixed by measuring the muon lifetime in the SM EFT,
\begin{align}
-\frac{4\hat{\mathcal{G}}_F}{\sqrt{2}} &=  -\frac{2}{\bar{v}_T^2} +  \left(C_{\substack{ll \\ \mu ee \mu}} +  C_{\substack{ll \\ e \mu\mu e}}\right) - 2 \left(C^{(3)}_{\substack{Hl \\ ee }} +  C^{(3)}_{\substack{Hl \\ \mu\mu }}\right).
\label{gfermi}
\end{align}
In the limit of $\rm{U(3)^5}$ flavour symmetry, this expression simplifies to
\bea\label{vtrelation}
\hat{G}_F = \frac{1}{\sqrt{2} \, \bar{v}_T^2} - \frac{1}{\sqrt{2}} \, C_{\substack{ll}} + \sqrt{2} \, C^{(3)}_{\substack{Hl}}.
\eea
We identify $\hat{\mathcal{G}}_F$ with the measured value of the Fermi constant in the  $\rm{U(3)^5}$ limit as $\hat{G}_F$ in this paper.
Our notation is such that a $1/\Lambda^2$ is implicit in each of the Wilson coefficients, and that $\bar{v}_T$ is the vev in the SMEFT given by
\bea
\bar{v}_T = \left(1 + \frac{3 \, C_H \, v^2}{8 \, \lambda} \right) v.
\eea
Here $\lambda$ is the coefficient of $(H^\dagger H)^2$ in the SM, with a normalization defined in the Appendix. $C_H$ is the Wilson coefficient of the $(H^\dagger H)^3$ operator, and $v$ is the SM vev in the limit $C_H \rightarrow 0$.
Many expressions that follow have explicit dependence on $\bar{v}_T$, which is related to $\hat{G}_F$ via Eqn \ref{vtrelation} as
\bea
\bar{v}_T^2 = \frac{1}{\sqrt{2} \hat{G}_F} + \frac{\delta G_F}{\hat{G}_F}, \quad {\rm when}, \quad \delta G_F =  \frac{1}{\sqrt{2} \,  \hat{G}_F} \left(\sqrt{2} \, C^{(3)}_{\substack{Hl}} - \frac{C_{\substack{ll}}}{\sqrt{2}}\right).
\eea
In what follows we use $ \delta G_F $, but note that the flavour dependence of this parameter is trivial to re-introduce,
and this shift can be considered to be implicitly flavour dependent.
\subsubsection{$M_Z$}
The mass eigenstate of the $Z$ boson is redefined as
\bea
\bar{M}_Z^2 = \frac{\bar{v}_T^2}{4}(\gcb^2+\gcw^2)+\frac{1}{8} \, \bar{v}_T^4 C_{HD} (\gcb^2+\gcw^2)+\frac{1}{2} \, \bar{v}_T^4 \gcb\gcw C_{HWB}.
\eea
The coupling of the $Z$ to fermions in the SM is proportional to $\sqrt{g_1^2+g_2^2}$. In the canonically normalized SMEFT, this coupling is
redefined and is proportional to  $\sqrt{\gcb^2+\gcw^2}$. As such it is convenient to define a $Z$ boson mass shift of the form
\bea\label{deltaMz}
\delta M_Z^2 \equiv  \frac{1}{2 \, \sqrt{2}} \, \frac{\hat{m}_Z^2}{\hat{G}_F} C_{HD} + \frac{2^{1/4} \sqrt{\pi} \, \sqrt{\hat{\alpha}} \, \hat{m}_Z}{\hat{G}_F^{3/2}} C_{HWB}.
\eea
This shift is useful in relating the effective couplings of the $Z$ to fermions in the SMEFT to the input parameters,
including $\hat{m}_Z$. The SM relations between Lagrangian parameters
and input parameters are used on the right hand side of Eqn \ref{deltaMz}, as the SMEFT corrections to these relations are higher order in $\bar{v}_T^2/\Lambda^2$.
As  $\hat{m}_Z$ is an input parameter, $\hat{m}_Z^2 = \bar{M}_Z^2$ in the SMEFT.

\subsubsection{$\sin^2 \theta$}
The kinetic mixing introduced by the operator with Wilson coefficient $C_{HWB}$ leads to a redefinition of the usual $s_\theta = \sin \theta$ mixing angle of the SM given by
\bea
s^2_\tc =  \frac{\gcb^2}{{\gcw^2 + \gcb^2}} + \frac{\gcb \gcw (\gcw^2-\gcb^2)}{(\gcb^2+\gcw^2)^2}  \bar{v}_T^2 C_{HWB}.
\eea
Here $s^2_\tc$ is used to rotate to the mass eigenstate fields in the SMEFT. As a short hand notation, we define
\bea
\delta s_\theta^2 \equiv \sin^2 \hat{\theta} -\sin^2 \bar{\theta} =  - \frac{s_\that \, c_\that}{2 \, \sqrt{2} \, \hat{G}_F (1 - 2 s^2_\that)} \left[s_\that \, c_\that \, (C_{HD} + 4 \, C^{(3)}_{\substack{H \ell}} - 2 \, C_{\substack{ll}})
+ 2 \, C_{HWB} \right].
\eea

\subsection{Gauge couplings in the SMEFT: $\bar{g}_1, \bar{g}_2$}
We relate the Lagrangian parameters $\bar{g}_2,\bar{g}_1$ to the input parameters at tree level via
\bea
\bar{g}_1^2  + \bar{g}_2^2  = 4 \, \sqrt{2} \, \hat{G}_F \, \hat{m}_Z^2 \left(1 - \sqrt{2} \,\delta G_F - \frac{\delta M_Z^2}{\hat{m}_Z^2} \right), \\
\bar{g}_2^2 = \frac{4 \, \pi \, \hat{\alpha}}{s^2_{\hat{\theta}}} \left[1 +  \frac{\delta s_\theta^2}{s^2_{\hat{\theta}}} + \frac{c_{\hat{\theta}}}{ s_{\hat{\theta}}} \frac{1}{\sqrt{2} \, \hat{G}_F} \, C_{HWB} \right].
\eea

\subsection{$M_W$ in the SMEFT}
The mass of the W boson is redefined in the SMEFT as
\bea
\bar{M}_W^2 = \frac{\bar{g}_{2}^2 \bar{v}_T^2}{4}.
\eea
Expressing $\bar{M}_W^2$ in terms of the inputs parameters we get:
\bea
\bar{M}_W^2 = \hat{m}_W^2 \left( 1 + \frac{\delta s_{\that}^2}{s_{\that}^2}+\frac{c_{\that}}{s_{\that} \sqrt{2} \hat{G}_F}C_{HWB} + \sqrt{2} \delta G_F \right) = \hat{m}_W^2 - \delta M_W^2,
\eea
where $\delta M_W^2= -\hat{m}_W^2 \left(\frac{\delta s_{\that}^2}{s_{\that}^2}+\frac{c_{\that}}{s_{\that} \sqrt{2} \hat{G}_F}C_{HWB} + \sqrt{2} \delta G_F\right)$
and $\hat{m}_W^2 = c_{\that}^2 \hat{m}_Z^2 $.

\section{Redefinition of Vector Boson couplings}\label{vectorcouplings}
\subsection{Neutral currents}
\subsubsection{Redefinition of $Z$ couplings}
The effective axial and vector couplings of the SMEFT $Z$ boson are defined as follows
\bea
\mathcal{L}_{Z,eff}  =  g_{Z,eff}  \,   \left(J_\mu^{Z \ell} Z^\mu + J_\mu^{Z \nu} Z^\mu + J_\mu^{Z u} Z^\mu +  J_\mu^{Z d} Z^\mu \right),
\eea
where $g_{Z,eff} = - \, 2 \, 2^{1/4} \, \sqrt{\hat{G}_F} \, \hat{m}_Z$, $(J_\mu^{Z x})^{pr} = \bar{x}_p \, \gamma_\mu \left[(\bar{g}^{x}_V)_{eff}^{pr}- (\bar{g}^{x}_A)_{eff}^{pr} \, \gamma_5 \right] x_r$ for $x = \{u,d,\ell,\nu \}$.
In general, these currents are matricies in flavour space. When we restrict our attention to the case of a minimal linear MFV scenario $(J_\mu^{Z x})_{pr} \simeq (J_\mu^{Z x}) \delta_{pr}$.
In the standard basis, the effective axial and vector couplings are modified from the SM values by a shift defined as
\bea
\delta (g^{x}_{V,A})_{pr} = (\bar{g}^{x}_{V,A})^{eff}_{pr} - (g^{x}_{V,A})^{SM}_{pr},
\eea
where
\bea\label{higherdgvga}
\delta (g^{\ell}_V)_{pr}&=&\delta \bar{g}_Z \, (g^{\ell}_{V})^{SM}_{pr} - \frac{1}{4 \sqrt{2} \hat{G}_F} \left(C_{\substack{H e \\pr}} + C_{\substack{H \ell \\ pr}}^{(1)} + C_{\substack{H \ell \\ pr}}^{(3)} \right) - \delta s_\theta^2, \\
\delta(g^{\ell}_A)_{pr}&=&\delta \bar{g}_Z \, (g^{\ell}_{A})^{SM}_{pr} + \frac{1}{4 \, \sqrt{2} \, \hat{G}_F}
\left( C_{\substack{H e \\pr}} - C_{\substack{H \ell \\ pr}}^{(1)} - C_{\substack{H \ell \\ pr}}^{(3)} \right),  \\
\delta (g^{\nu}_V)_{pr}&=&\delta \bar{g}_Z \, (g^{\nu}_{V})^{SM}_{pr} - \frac{1}{4 \, \sqrt{2} \, \hat{G}_F} \left( C_{\substack{H \ell \\ pr}}^{(1)} - C_{\substack{H \ell \\ pr}}^{(3)} \right),
\\
\delta(g^{\nu}_A)_{pr}&=&\delta \bar{g}_Z \,(g^{\nu}_{A})^{SM}_{pr}  - \frac{1}{4 \, \sqrt{2} \, \hat{G}_F}
\left(C_{\substack{H \ell \\ pr}}^{(1)} - C_{\substack{H \ell \\ pr}}^{(3)} \right),  \\
\delta (g^{u}_V)_{pr}&=&\delta \bar{g}_Z \, (g^{u}_{V})^{SM}_{pr}  +
\frac{1}{4 \, \sqrt{2} \, \hat{G}_F} \left(- C_{\substack{H q \\ pr}}^{(1)} + \, C_{\substack{H q \\ pr}}^{(3)} -C_{\substack{H u \\ pr}} \right) + \frac{2}{3} \delta s_\theta^2,\\
\delta(g^{u}_A)_{pr}&=&\delta \bar{g}_Z \, (g^{u}_{A})^{SM}_{pr}
-\frac{1}{4 \, \sqrt{2} \, \hat{G}_F} \left( C_{\substack{H q \\ pr}}^{(1)} -  \, C_{\substack{H q \\ pr}}^{(3)} - C_{\substack{H u \\ pr}} \right),  \\
\delta (g^{d}_V)_{pr}&=&\delta \bar{g}_Z \,(g^{d}_{V})^{SM}_{pr}
-\frac{1}{4 \, \sqrt{2} \, \hat{G}_F} \left( C_{\substack{H q \\ pr}}^{(1)}  +  \, C_{\substack{H q \\ pr}}^{(3)} + C_{\substack{H d \\ pr}} \right) -  \frac{1}{3} \delta s_\theta^2, \\
\delta(g^{d}_A)_{pr}&=&\delta \bar{g}_Z \,(g^{d}_{A})^{SM}_{pr}
+\frac{1}{4 \, \sqrt{2} \, \hat{G}_F} \left(-C_{\substack{H q \\ pr}}^{(1)}  -  \, C_{\substack{H q \\ pr}}^{(3)} + C_{\substack{H d \\ pr}} \right).
\eea
where
\bea
\delta \bar{g}_Z =- \frac{\delta G_F}{\sqrt{2}} - \frac{\delta M_Z^2}{2\hat{m}_Z^2} + \frac{s_{\hat{\theta}} \, c_{\hat{\theta}}}{\sqrt{2} \hat{G}_F} \, C_{HWB}
\eea
Here our chosen normalization is $(g^{x}_{V})^{SM} = T_3/2 - Q^x \, \bar{s}_\theta^2, (g^{x}_{A})^{SM} = T_3/2$ where $T_3 = 1/2$ for $u_i,\nu_i$ and $T_3 = -1/2$ for $d_i,\ell_i$
and $Q^x = \{-1,2/3,-1/3 \}$ for $x = \{\ell,u,d\}$.

\subsubsection{Redefinition of $A$ couplings}
For the electromagnetic current we define:
\bea
\mathcal{L}_{A,eff}=-\sqrt{4 \pi \hat{\alpha}}  \left[ Q_{x} \, J_{\mu}^{A, x}  \right] A^{\mu} .
\eea
for $x = \ell,u,d$. The measured effective electromagnetic coupling $\hat{\alpha}$ is directly identified
with the modified coupling present in the SMEFT: $\bar{\alpha} = \bar{e}^2/4 \pi$, with $\bar{e}$ given by
\bea
\bar{e} = \bar{g}_2 \, s_{\bar{\theta}}  \left[1 -  \frac{c_\that}{ \, s_\that} \, \frac{1}{2 \, \sqrt{2} \hat{G_F}} \, C_{HWB}\right]= \sqrt{4 \pi \hat{\alpha}}.
\eea
This means the shift in the definition of $\alpha$ given in the previous equation is unobservable, considering our chosen input parameters. As such we can trade
$\bar{\alpha}$ directly for $\hat{\alpha}$.

\subsection{Charged currents}

For the charged currents, we define
\bea
\mathcal{L}_{W,eff} = - \frac{ \sqrt{2 \,\pi \, \hat{\alpha}}}{s_\that} \left[(J_{\mu}^{W_\pm, \ell})_{pr} W_\pm^\mu + (J_{\mu}^{W_\pm, q})_{pr} W_\pm^\mu\right],
\eea
where in the SM one has
\bea
(J_\mu^{W_{+}, \ell})_{pr} &=&   \, \bar{\nu}_p \, \gamma^\mu \,\left(\bar{g}^{W_{+},\ell}_V - \bar{g}^{W_{+},\ell}_A \gamma_5 \right)\, \ell_r, \\
(J_\mu^{W_{-}, \ell})_{pr} &=& \, \bar{\ell}_p \, \gamma^\mu \, \left(\bar{g}^{W_{-},\ell}_V - \bar{g}^{W_{-},\ell}_A \gamma_5 \, \right) \nu_r.
\eea
In the SMEFT we note that in the flavour symmetric limit
\bea
\delta(g^{W_{\pm},\ell}_V)_{rr} = \delta(g^{W_{\pm},\ell}_A)_{rr}  &=&  \frac{1}{2\sqrt{2} \hat{G}_F} \left(C^{(3)}_{\substack{H \ell \\ rr}} + \frac{1}{2} \frac{c_{\hat{\theta}}}{ s_{\hat{\theta}}} \, C_{HWB} \right)
+ \frac{1}{4} \frac{\delta s_\theta^2}{s^2_{\hat{\theta}}}.
\eea

Note that although the corrections in the SMEFT shown preserve the left handed structure of the current for the lepton couplings, we introduce a separate
axial and vector coupling for later convenience. For the quark charged currents one similarly finds
\bea
\delta(g^{W_{\pm},q}_V)_{rr} = \delta(g^{W_{\pm},q}_A)_{rr}  &=&  \frac{1}{2\sqrt{2} \hat{G}_F} \left(C^{(3)}_{\substack{H q \\ rr}} + \frac{1}{2} \frac{c_{\hat{\theta}}}{ s_{\hat{\theta}}} \, C_{HWB} \right)
+ \frac{1}{4} \frac{\delta s_\theta^2}{s^2_{\hat{\theta}}}.
\eea
There is also dependence on the operator $Q_{\substack{H ud \\ rr}}$ for the $W$ quark current.
When we assume linear MFV,
the Wilson coefficient of this operator is suppressed by
\bea
C_{\substack{H ud \\ rr}}  \propto \left[Y_u \, Y_d^\dagger \right]_{rr},
\eea
and in this case, this contribution is neglected for reasons of consistency.
Light quark mass suppressed corrections are neglected in the SM predictions of many of the observables considered here,
and also when higher dimensional operators are inserted.

\section{Observables}\label{obs}
Whenever possible, we express all observables in terms of shifts of the form
\bea
\delta G_F, \, \,  \delta M_Z^2, \, \, \delta M_W^2, \, \,   \delta s^2_\theta, \, \,  \delta g^{x}_{V,A}, \, \, \delta g^{W_\pm, y}_{V.A}.
\eea
Here $x = \ell, u,d$ and $y = \ell,q$. Added to these corrections for each observable are contributions due to explicit operator insertions that are not (easily) expressible in terms of these common shifts.
These net shift variables do not correspond to a basis for $\lsix$, they are simply a convenient shorthand notation for some terms in the effective Lagrangian.

\subsection{Differential cross section for $\ell^{+} \ell^{-} \rightarrow f \bar{f}$}\label{differential}

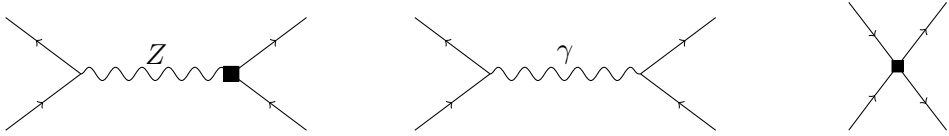
\begin{figure}
\hspace{0.6cm}
\begin{tikzpicture}
\draw[decorate,decoration=snake] (-0.5,0) -- (1.5,0) ;
\filldraw (1.4,-0.1) rectangle (1.6,0.1);
\draw [->](1.5,0) -- (2.1,0.45);
\draw (2.1,0.45) -- (2.5,0.75);
\draw (1.5,0) -- (2,-0.375);
\draw [<-](2,-0.375) -- (2.5,-0.75);

\draw [->](-0.5,0) -- (-1.1,0.45);
\draw (-1.1,0.45) -- (-1.5,0.75);
\draw (-0.5,0) -- (-1,-0.375);
\draw [<-](-1,-0.375) -- (-1.5,-0.75);

\node [above] at (0.5,0) {$Z$};

\end{tikzpicture}\hspace{1.3cm}
\begin{tikzpicture}
\draw[decorate,decoration=snake] (-0.5,0) -- (1.5,0) ;
\draw [->](1.5,0) -- (2.1,0.45);
\draw (2.1,0.45) -- (2.5,0.75);
\draw (1.5,0) -- (2,-0.375);
\draw [<-](2,-0.375) -- (2.5,-0.75);

\draw [->](-0.5,0) -- (-1.1,0.45);
\draw (-1.1,0.45) -- (-1.5,0.75);
\draw (-0.5,0) -- (-1,-0.375);
\draw [<-](-1,-0.375) -- (-1.5,-0.75);

\node [above] at (0.5,0) {$\gamma$};
\end{tikzpicture}
\hspace{1.5cm}
\begin{tikzpicture}
\draw[->] (-0.65,-0.85) -- (-0.3,-0.385);
\draw [->](-0.3,-0.385) -- (0.38,0.5) ;
\draw (0.38, 0.5) --(0.65,0.85);
\draw [->](-0.65,0.85) -- (-0.3,0.395);
\draw [->](-0.3, 0.395) -- (0.38,-0.5) ;
\draw  (0.38,-0.5)--(0.65,-0.85);
\filldraw (-0.075,-0.075) rectangle (0.075,0.075);
\end{tikzpicture}
\caption{\label{fig:1}
Diagrams contributing to near $Z$ pole $2 \rightarrow 2$ scattering in the SMEFT. The black box indicates the insertion of $\lsix$.}
\end{figure}

Observables that are not limited to the $Z$ pole are an important source of information on
Wilson coefficients present in the SMEFT.  Corrections to the $2 \rightarrow 2$ differential spectrum
predicts the total cross sections $\sigma_{\ell^{+} \ell^{-} \rightarrow f \, \bar{f}}$ where $f=\{\ell,u,c,b,d,s\}$ (here the final and initial state leptons are defined to not have the same flavour), as well as the differential and angular observables
for these processes. A general expression in the SMEFT valid for on and off resonance scattering includes a contribution from $Z$ and $\gamma$ exchange as well as the effect of
$\psi^4$ operators and the interference of all of these terms, see Fig \ref{fig:1}.
Our discussion of this general expression in the SMEFT will largely build on the discussion in Ref \cite{Z-Pole}  which itself borrows heavily from Ref \cite{Bardin:1999ak}.\footnote{For classic related results, that are outside of the systematic SMEFT analysis presented here,
see Ref \cite{Eichten:1983hw}.}

Up to leading order in the interference of the $\psi^4$ operators with the SM contributions, the general differential expression for $\ell^- \, \ell^+ \rightarrow f \, \bar{f}$ is as follows.
Here we neglect initial and final state radiation (including possible $\alpha_s$ corrections to final state fermions), initial and final state fermion masses are neglected, and the initial $e^+,e^-$ are assumed to be unpolarized. The general $s$ channel expression we find for the SMEFT is\footnote{In this expression we have used Feynman gauge.}

\bea\label{generaldifferential}
\frac{1}{N_c}\frac{d\sigma}{d c_\theta} &=&  \frac{\hat{G}_F^2 \hat{m}_Z^4}{\pi}
\, \bar{\chi}(s) \left[\left(|\bar{g}^{\ell}_V|^2 + |\bar{g}^{\ell}_A|^2\right) \, \left(|\bar{g}^{f}_V|^2 + |\bar{g}^{f}_A|^2\right) \left(1+ c_\theta^2 \right)
- 8 \, {\rm Re}\left[\bar{g}^{\ell}_A  \bar{g}^{\ell,\star}_V \right] \, {\rm Re}\left[\bar{g}^{f}_A \bar{g}^{f, \star}_V \right] c_\theta \right], \nn
&+& \frac{|\hat{\alpha}|^2 \, |Q_\ell|^2 \, |Q_f|^2 \, \pi}{2 \, s} \left(1+ c_\theta^2 \right) + \frac{\hat{G}_F \hat{m}_Z^2 Q_\ell \, Q_f}{\sqrt{2}} \,
\left[\alpha^\star \frac{\bar{g}^{\ell}_V \, \bar{g}^{f}_V \left(1+ c_\theta^2 \right) + 2 \, c_\theta \, \bar{g}^{\ell}_A \, \bar{g}^{f}_A}{s - \bar{M}_Z^2 + i \,  \bar{w}(s)} + {\rm h.c.} \right], \nn
&+& \frac{Q_\ell \, Q_f}{32} \, \left[\alpha^\star \, C_{LL,RR}^{\ell, f} \, (1+ c_\theta)^2  + {\rm h.c.}  \right] +
\frac{Q_\ell \, Q_f}{32} \, \left[\alpha^\star \, C_{LR}^{\ell, f} \, (1- c_\theta)^2  + {\rm h.c.}  \right], \\
&+&  \left(\frac{\hat{G}_F \hat{m}_Z^2}{16 \, \sqrt{2} \, \pi} \right)\left[ \left(\frac{s }{s - \bar{M}_Z^2 + i \, \bar{w}(s)} \right)
C_{LL,RR,LR}^{\ell, f,\star} (\bar{g}^{\ell}_V \pm \bar{g}^{\ell}_A)(\bar{g}^{f}_V \pm \bar{g}^{f}_A) \left(1+ c_\theta^2 \right) + {\rm h.c.} \right], \nn
&+&  \left(\frac{\hat{G}_F \hat{m}_Z^2}{16 \, \sqrt{2} \, \pi} \right) \left[ \left(\frac{s}{s - \bar{M}_Z^2 + i \, \bar{w}(s)}\right)
 C_{LL,RR,LR}^{\ell, f,\star} \, (\bar{g}^{\ell}_A \pm \bar{g}^{\ell}_V)(\bar{g}^{f}_A \pm \bar{g}^{f}_V) \, 2 \, c_\theta + {\rm h.c.}\right]. \nonumber
\eea

We have used in the expression
\bea
\bar{\chi}(s) = \frac{s}{(s - \bar{M}_Z^2)^2 + | \bar{w}(s)|^2}.
\eea
The Breit-Wigner distribution \cite{Breit:1936zzb} is introduced as $\bar{w}(s)$, and we treat this as a possibly $s$ dependent function to maintain generality.
A possible choice for the Breit-Wigner distribution  is the use of an $s$ dependent width ($\bar{w}(s) = s \, \bar{\Gamma}_Z /\bar{M}_Z$), which is the approach used at LEP, as discussed in Ref \cite{Z-Pole,Altarelli:1989hv}. Alternatively the real part of the complex pole can be directly used
introducing $\bar{w}(s) =\bar{\Gamma}_Z \,  \bar{M}_Z$ for the Breit-Wigner distribution.
These prescriptions can be mapped to one another in the SM, see Ref \cite{Sirlin:1991fd}.
The latter pole specification is strongly preferred in our view, we simply introduce $\bar{w}(s)$ to remain as general as possible
as a notation convention.

Four fermion operators that interfere and contribute are denoted $C_{LL,RR,LR}$, and  are in the classes $LL,RR$ and $LR$ for the operator basis specified in Ref. \cite{Grzadkowski:2010es}.
In Eqn. \ref{generaldifferential} the $+/-$ expressions for the $\ell,f$ couplings correspond to the case of the $L/R$ projectors present in the
$\psi^4$ operators respectively.
In Eqn. \ref{generaldifferential} we have suppressed flavour indicies on the $\psi^4$ operator Wilson coefficients and the effective gauge couplings.
Reintroducing the flavour indicies on the $\psi^4$ operators, one finds $C^\star \rightarrow C^\star_{\ell \, \ell \, f \, f}, C^\star_{\ell \, f \, f \, \ell}, C^\star_{f \, \ell \, \ell \, f}$ for $C^\star_{LL,RR}$.
For the $LR$ operators $C^\star \rightarrow C^\star_{\ell \, \ell \, f \, f}$ is as in the previous chirality cases, while the cases $ C^\star_{\ell \, f \, f \, \ell},C^\star_{f \, \ell \, \ell \, f}$ vanish.

The parameter $c_\theta$ is the angle between the incoming $\ell^-$ and the outgoing $\bar{f}$, and $s = (p_{\ell^+} + p_{\ell^-})^2$. $N_C$ is the dimension of the $\rm SU(3)$ group
of the produced fermion $f$. Note that $\alpha$ can obtain a small imaginary contribution in the running of this coupling. The theoretical prediction of this expression also depends on $\bar{M}_Z, \bar{\Gamma}_Z,\bar{g}^{\ell,f}_{A,V}$ which are the theoretical effective mass, width
and couplings in the SMEFT.

When considering $\ell^- \, \ell^+ \rightarrow \ell^- \, \ell^+$ for differential and total cross section observables,
$t$ channel contributions are also present, and the interference effects of the $\psi^4$ operators are modified. We restrict our attention initially to
$\ell^- \, \ell^+ \rightarrow f \, \bar{f}$ where $f$ is defined to not be the same state as the initial state fermion. The case when all of the initial and final states are the same
fermion is discussed in Section \ref{samestate}

In Eqn. \ref{generaldifferential} we have neglected interference effects with operators of the form $LRRL, LRLR$ that are
proportional to SM Yukawas (and hence light quark masses) in the case of $\rm U(3)^5$ symmetry being assumed in the SMEFT.

\subsubsection{Scaling of SMEFT corrections}\label{scaling}

The scaling of the corrections on and off the $Z$ pole is of interest. {\it Near} the $Z$ pole, the contributions due to $\lsix$ interfering with the SM in
Eqn. \ref{generaldifferential} have the general scaling:
\bea\label{scaling}
&Z-Z&: \sim  \frac{\bar{v}_T^2}{\Gamma_Z^2 \, \Lambda^2}, \quad \gamma - \gamma: \sim \frac{\bar{v}_T^2}{M_Z^2 \, \Lambda^2}, \quad Z-\gamma:  \sim \frac{\bar{v}_T^2}{M_Z^2 \, \Lambda^2},
\nn
&\psi^4& \! \! \! \! \! \! -Z: \sim \frac{1}{\Lambda^2}, \quad \psi^4 - \gamma: \sim \frac{1}{\Lambda^2}.
\eea
Here $Z,\gamma$ corresponds to a Gauge boson exchange and $\psi^4$ corresponds to a four fermion operator in $\lsix$.
A few comments are in order considering these estimates. The usual choices of Breit-Wigner distribution used in Eqn \ref{generaldifferential} do not change these scaling estimates.
Exactly on the $Z$ pole the interference due to $\gamma -Z$ and $Z -\psi^4$ contributions vanish. A large fraction of LEPI data
is taken at $\sqrt{s} - M_Z  \sim \Gamma_Z$, where these sub-leading terms scale as in Eqn \ref{scaling}.  The combined LEPI data set analysis, with on and off pole $Z$ data, determines EWPD parameters. It is tempting to conclude that the subdominant contributions can be completely neglected for near $Z$ pole data as $\Gamma_Z^2/\bar{v}_T^2 \sim \mathcal{O}(10^{-3})$. However, the scaling of these suppressed contributions in the partial widths extracted from LEPI data is relatively suppressed by $\Gamma_Z \, M_Z/\bar{v}_T^2$
compared to the leading effect of dimension six operators, as we will show.

Further, for measurements at LEPII taken at $\sqrt{s} \sim 2 M_z$, these corrections have the scaling
 \bea
&Z-Z&: \sim \frac{\bar{v}_T^2}{M_Z^2 \, \Lambda^2}, \quad \gamma - \gamma: \sim \frac{\bar{v}_T^2}{M_Z^2 \, \Lambda^2}, \quad Z-\gamma:  \sim \frac{\bar{v}_T^2}{M^2_Z \, \Lambda^2},
\nn
&\psi^4& \! \! \! \! \! \! -Z: \sim \frac{1}{\Lambda^2}, \quad \psi^4 - \gamma: \sim \frac{1}{\Lambda^2}.
\eea
In these measurements the subdominant contributions of $\psi^4$ operators are only suppressed by $M_Z^2/\bar{v}_T^2$ and must be included.
At the LHC, the EW process $\bar{f} \, f \rightarrow \ell^- \, \ell^+$ is potentially accessible at larger $s$. Assuming $s \gg M_Z^2$ one has the scaling
 \bea
&Z-Z&: \sim \frac{\bar{v}_T^2}{s \, \Lambda^2}, \quad \gamma - \gamma: \sim \frac{\bar{v}_T^2}{s \, \Lambda^2}, \quad Z-\gamma:  \sim \frac{\bar{v}_T^2}{s \, \Lambda^2},
\nn
&\psi^4& \! \! \! \! \! \! -Z: \sim \frac{1}{\Lambda^2}, \quad \psi^4 - \gamma: \sim \frac{1}{\Lambda^2}.
\eea

The assumption that $s \ll \Lambda^2$ is implicit, but can be challenged,
particularly for larger $s$ measurements at LHC. When the expansion in local operators breaks down, the operators can be resumed into
effective form factors\footnote{See Refs \cite{Isidori:2013cla,Isidori:2013cga,Gonzalez-Alonso:2014eva} for some discussion.} which can be extracted from differential distributions, or at fixed $s$.
These simple scaling estimates neglect order one factors, but make clear the requirement that a global analysis including LEPII data and LHC data include
these corrections when precise (and accurate) bounds are of interest in the SMEFT.\footnote{These subdominant corrections are also suppressed
by some function of the off pole data in the total data set, compared to the data taken exactly on the pole. The most naive such scaling yields a factor of $\sim 2/10$.}

\subsubsection{$\psi^4$ operators and $\rm U(3)^5$}

The $\psi^4$ operators that can contribute significantly to offshell  $\ell^{+} \ell^{-} \rightarrow f \bar{f}$ and $\ell^{+} \ell^{-} \rightarrow \ell^{+} \ell^{-}$ are
\bea
\mathcal{L}_{\psi^4} &=&  C_{\substack {\ell \, \ell \\ p r s t }} \, Q_{\substack {\ell \, \ell \\ p r s t }} +
C^{(1)}_{\substack {\ell \, q \\ p r s t }} \, Q^{(1)}_{\substack {\ell \, q \\ p r s t }} +
C^{(3)}_{\substack {\ell \, q \\ p r s t }} \, Q^{(3)}_{\substack {\ell \, q \\ p r s t }}  +
C_{\substack {e \, e \\ p r s t }} \, Q_{\substack {e \, e \\ p r s t }} +
C_{\substack {e \, u \\ p r s t }} \, Q_{\substack {e \, u \\ p r s t }} +
C_{\substack {e \, d \\ p r s t }} \, Q_{\substack {e \, d \\ p r s t }}, \nn
&+&
C_{\substack {\ell \, e \\ p r s t }} \, Q_{\substack {\ell \, e \\ p r s t }}+
C_{\substack {\ell \, u \\ p r s t }} \, Q_{\substack {\ell \, u \\ p r s t }} +
C_{\substack {\ell \, d \\ p r s t }} \, Q_{\substack {\ell \, d \\ p r s t }} +
C_{\substack {q \, e \\ p r s t }} \, Q_{\substack {q \, e \\ p r s t }} +
C^{(1)}_{\substack {\ell \, e \, q \, u \\ p r s t }} \, Q^{(1)}_{\substack {\ell \, e \, q \, u \\ p r s t }}, \nn
&+& C^{(3)}_{\substack {\ell \, e \, q \, u \\ p r s t }} \, Q^{(3)}_{\substack {\ell \, e \, q \, u \\ p r s t }}
+ C_{\substack {\ell \, e \, d \, q \\ p r s t }} \, Q_{\substack {\ell \, e \, d \, q \\ p r s t }}.
\eea
These operators are in general not Hermitian in flavour space and can have complex Wilson coefficients.
Nevertheless the interference effect of the operators with the SM tree level processes vanishes for the complex part of the Wilson coefficients,
as there are no flavour changing neutral currents at tree level in the SM.

Of these operators, the following are not suppressed by the insertion of light fermion masses when $\rm U(3)^5$ is
assumed
\bea\label{nonchiral}
Q_{\substack {\ell \, \ell \\ p r s t }} &= (\overline \ell_p \gamma_\mu \ell_r)(\overline \ell_s \gamma^\mu \ell_t), \quad \quad &
Q^{(1)}_{\substack {\ell \, q \\ p r s t }} = (\overline \ell_p \gamma_\mu \ell_r)(\overline q_s \gamma^\mu q_t), \\
Q^{(3)}_{\substack {\ell \, q \\ p r s t }} &= (\overline \ell_p \gamma_\mu \tau_i \ell_r)(\overline q_s \gamma^\mu \tau_i q_t), \quad \quad &
Q_{\substack {e e \\ p r s t }} = (\overline e_p \gamma_\mu e_r)(\overline e_s \gamma^\mu e_t),\\
Q_{\substack {e \, u \\ p r s t }} &= (\overline e_p \gamma_\mu e_r)(\overline u_s \gamma^\mu u_t), \quad \quad &
Q_{\substack {e d \\ p r s t }} = (\overline e_p \gamma_\mu e_r)(\overline d_s \gamma^\mu d_t),\\
Q_{\substack {\ell \, e \\ p r s t }} &= (\overline \ell_p \gamma_\mu \ell_r)(\overline e_s \gamma^\mu e_t),  \quad \quad &
Q_{\substack {\ell u \\ p r s t }} = (\overline \ell_p \gamma_\mu \ell_r)(\overline u_s \gamma^\mu u_t),\\
Q_{\substack {\ell \, d \\ p r s t }} &= (\overline \ell_p \gamma_\mu \ell_r)(\overline d_s \gamma^\mu d_t),  \quad \quad &
Q_{\substack {q \, e \\ p r s t }} = (\overline q_p \gamma_\mu q_r)(\overline e_s \gamma^\mu e_t).
\label{nonchiralf}
\eea
When $\rm U(3)^5$ symmetry is assumed for $\lsix$,
the Wilson coefficients of the operators in Eqn \ref{nonchiral}-\ref{nonchiralf} are all proportional to $ \delta_{pr} \, \delta_{st}$.
These operators add three unknown parameters into constraints obtained from purely leptonic EWPD far off the $Z$ pole.
Precision electroweak data involving final state up quarks depends on
four extra parameters off the $Z$ pole due to these $\psi^4$ operators, as does precision data involving final state down quarks. Two of these parameters
(due to $C^{(1)}_{\substack {\ell \, q \\ p p s s }}$ and $C_{\substack {q \, e \\ p p s s }}$) are common for the final state quark
cases.

The remaining $\psi^4$ operators that are proportional to light quark masses (in a $\rm U(3)^5$ scenario) are
\bea
Q^{(1)}_{\substack {\ell \, e \, q \, u \\ p r s t }} &= (\overline \ell^i_p  e_r) \epsilon_{ij} (\overline q^j_s  u_t), \quad \quad &
Q^{(3)}_{\substack {\ell \, e \, q \, u \\ p r s t }} =  (\overline \ell^i_p  \sigma_{\mu \, \nu} \, e_r) \epsilon_{ij} (\overline q^j_s \, \sigma^{\mu \, \nu}\,   u_t), \\
Q_{\substack {\ell \, e \, d \, q \\ p r s t }} & \! \! \! \! \! \! \! \! \!  = (\overline \ell^i_p  e_r) (\overline d_s  q_{t,i}).
\eea

\subsubsection{Shifts in differential $\ell^{+} \ell^{-} \rightarrow f \bar{f}$ spectra}
The shift in the differential $\ell^{+} \ell^{-} \rightarrow f \bar{f}$ spectra in the SMEFT $\frac{1}{N_c} \delta \left(\frac{d\sigma}{d c_\theta}\right)$ is given by:
\bea\label{generaldifferentialshift}
&\,& \frac{\hat{G}_F^2 \hat{m}_Z^4}{\pi}
\, \chi(s) \left[2  {\rm Re}\left[G^{\ell *}_V \, \delta g^{\ell}_V +G^{\ell *}_A \, \delta g^{\ell}_A\right] \, \left(|G^{f}_V|^2 + |G^{f}_A|^2\right) \left(1+ c_\theta^2 \right)
+  \left(\ell \leftrightarrow f \right) \right], \\
&-& \frac{8}{\pi} \hat{G}_F^2 \hat{m}_Z^4 \, \chi(s) \left[
 {\rm Re}\left[\delta g^{\ell}_A  G^{\ell,\star}_V + G^{\ell}_A  \delta g^{\ell,\star}_V \right] \, {\rm Re}\left[G^{f}_A G^{f, \star}_V \right] c_\theta +  \left(\ell \leftrightarrow f \right) \right], \nn
 &+& \frac{\hat{G}_F^2 \hat{m}_Z^4}{\pi}
\, \delta \chi(s) \left[\left(|G^{\ell}_V|^2 + |G^{\ell}_A|^2\right) \, \left(|G^{f}_V|^2 + |G^{f}_A|^2\right) \left(1+ c_\theta^2 \right)
- 8 \, {\rm Re}\left[G^{\ell}_A  G^{\ell,\star}_V \right] \, {\rm Re}\left[G^{f}_A  G^{f, \star}_V \right] c_\theta \right], \nn
 &+& \frac{\hat{G}_F \hat{m}_Z^2 Q_\ell \, Q_f}{\sqrt{2}} \,
\left[\alpha^\star  \chi_2(s) \, \frac{(\delta g^{\ell}_V \, G^{f}_V+  G^{\ell}_V \, \delta g^{f}_V) \left(1+ c_\theta^2 \right) + 2 \, c_\theta \,
\left(\delta g^{\ell}_A \, G^{f}_A +  G^{\ell}_A \, \delta g^{f}_A\right) }{s} + {\rm h.c.} \right], \nn
&+& \frac{\hat{G}_F \hat{m}_Z^2 Q_\ell \, Q_f}{\sqrt{2}} \,
\left[\alpha^\star  \delta \chi_2(s) \, \frac{G^{\ell}_V \, G^{f}_V \left(1+ c_\theta^2 \right) + 2 \, c_\theta \,
G^{\ell}_A \, G^{f}_A}{s} + {\rm h.c.} \right], \nn
&+& \frac{Q_\ell \, Q_f}{32} \, \left[\alpha^\star \, C_{LL,RR}^{\ell, f} \, (1+ c_\theta)^2  + {\rm h.c.}  \right] +
\frac{Q_\ell \, Q_f}{32} \, \left[\alpha^\star \, C_{LR}^{\ell, f} \, (1- c_\theta)^2  + {\rm h.c.}  \right], \nn
&+&  \left(\frac{\hat{G}_F \hat{m}_Z^2}{16 \, \sqrt{2} \, \pi} \right)\left[ \chi_2(s)
C_{LL,RR,LR}^{\ell, f,\star} (G^{\ell}_V \pm G^{\ell}_A)(G^{f}_V \pm G^{f}_A) \left(1+ c_\theta^2 \right) + {\rm h.c.} \right], \nn
&+&  \left(\frac{\hat{G}_F \hat{m}_Z^2}{16 \, \sqrt{2} \, \pi} \right) \left[  \chi_2(s) C_{LL,RR,LR}^{\ell, f,\star} \, (G^{\ell}_A \pm G^{\ell}_V)(G^{f}_A \pm G^{f}_V) \, 2 \, c_\theta + {\rm h.c.}\right]. \nonumber
\eea
Here we have introduced the notation $G^{\ell,f}_{A,V}$ which corresponds to the leading order prediction of an $Z$ axial or vector coupling
in the SM, for the state $\ell,f$. We have also introduced
\begin{align}
\chi(s) &= |\Xi(s)|^2/s, \quad   \quad  &\delta \chi(s) &= \frac{1}{s} \left[\Xi(s) \,  \delta \Xi^\star(s) + \delta \Xi(s) \, \Xi^\star(s)\right],  \\
\chi_2(s) &=\Xi(s), \quad \quad & \delta \chi_2(s) &= \delta \, \Xi(s),
\end{align}
where the $\left( \cdots \right)_{SM}$ expressions are defined to be the leading order SM theoretical predictions of the quantities in the parenthesis
and
\bea
\Xi(s) &=& \frac{s}{s -  \hat{m}_Z^2 + i (w(s))_{SM} }, \\
\delta \Xi(s) &=& \frac{s}{[s -  \hat{m}_Z^2 + i (w(s))_{SM}]^2} \left[ - i \delta w(s) \right]. \\
\text{With : } \bar{w}(s) &=& s \frac{\bar{\Gamma}_Z}{\bar{M}_Z} \text{ we get : } \delta w(s) = s \left(\frac{(\Gamma_Z)_{SM}}{\hat{m}_Z}\right)\left( \frac{\delta \Gamma_Z}{(\Gamma_Z)_{SM}}\right). \\
\text{With : } \bar{w}(s) &=&  \bar{\Gamma}_Z \bar{M}_Z \text{ we get : } \delta w(s) =  (\Gamma_Z)_{SM} \hat{m}_Z\left( \frac{\delta \Gamma_Z}{(\Gamma_Z)_{SM}}\right).
\eea

\subsubsection{Differential cross section for $\bar{F} \, F \rightarrow \bar{F} \, F$}\label{samestate}
The case $\bar{F} \, F \rightarrow \bar{F} \, F$ where $F$ is a fermion and the initial and final states are identical
has two kinematic channels, s and t, present. Of particular interest considering LEP data,
is the case $\bar{\ell} \, \ell \rightarrow \bar{\ell} \, \ell$ where $\ell = e$. Adopting the same set of approximations and assumptions
as in Section \ref{differential}, Bhabba scattering ($e^+\ e^- \rightarrow e^+\ e^-$) in the SMEFT is given by
\bea
 \frac{d \sigma}{d c_\theta} &=& \frac{2 \, \hat{G}_F^2 \hat{m}_Z^{4}}{\pi s}\left[
( |\bar{g}^{\ell}_V|^2 + |\bar{g}^{\ell}_A|^2 )^2 \left(\frac{u^2 + s^2}{(t-\bar{M}_Z^2)^2} +  \frac{\bar{\chi}(s)}{s} \left(u^2 + t^2 \right) +  2 \, \bar{\chi}(s) \frac{u^2 (1 - \bar{M}_Z^2/s)}{t - \bar{M}_Z^2} \right), \right. \nn
&\,& \left. \hspace{1.75cm} - 4 \, {\rm Re}\left[\bar{g}^{\ell *}_V \bar{g}^{\ell}_A\right]^2 \left(\frac{s^2 - u^2}{(t-\bar{M}_Z^2)^2} +  \frac{\bar{\chi}(s)}{s} \left(u^2 - t^2 \right) - 2 \, \bar{\chi}(s) \frac{u^2 (1 - \bar{M}_Z^2/s)}{t - \bar{M}_Z^2} \right) \right], \nn
&+& \frac{ \sqrt{2} \, \hat{G}_F \hat{m}_Z^2}{s} \left[\hat{\alpha}^* \frac{(\bar{g}^{\ell}_V)^2 (u^2 + t^2) + (\bar{g}^{\ell}_A)^2 (u^2-t^2)}{s \left(s - \bar{M}_Z^2 + i \bar{w}(s) \right)} + \hat{\alpha}^* \frac{(\bar{g}^{\ell}_V)^2 (u^2 + s^2) + (\bar{g}^{\ell}_A)^2 (u^2-s^2) }{t \left(t - \bar{M}_Z^2 \right)} + h.c. \right], \nn
 &+& \frac{\sqrt{2} \, \hat{G}_F \hat{m}_Z^2 \, u^2}{s}\left[\frac{\hat{\alpha}^*}{t}
 \frac{(\bar{g}^{\ell}_V)^2 + (\bar{g}^{\ell}_A)^2}{\left(s - \bar{M}_Z^2 + i \bar{w}(s) \right)} +
  \frac{\hat{\alpha}}{s} \frac{(\bar{g}^{\ell,\star}_V)^2 + (\bar{g}^{\ell,\star}_A)^2}{\left(t - \bar{M}_Z^2 \right)}  \right], \nonumber \\
 &+&  \frac{2 \, \pi \, \hat{\alpha}^{2}}{s}\left[\frac{u^2 + s^2}{t^2}+ \frac{u^2+t^2}{s^2} + \frac{2 u^2}{ts}\right]
 + \frac{\hat{\alpha}}{4 s}\left[ 2\left(\frac{u^2}{s} + \frac{u^2}{t} \right)C_{LL,RR}^\star + \left(\frac{t^2}{s}+ \frac{s^2}{t} \right)C^\star_{LR} + h.c \right], \nonumber\\
 &+& \frac{\hat{G}_F \hat{m}_Z^2}{4 \sqrt{2} \pi s} \left[\frac{4 u^2 \left(\bar{g}^{\ell}_A \pm \bar{g}^{\ell}_V\right)^2 C^\star_{LL,RR}  + 2 t^2\left((\bar{g}^{\ell}_V)^2 - (\bar{g}^{\ell}_A)^2\right)C^\star_{LR}}{s-\bar{M}_Z^2 +i w(s)} + h.c \right], \nonumber \\
  &+& \frac{\hat{G}_F \hat{m}_Z^2}{4 \sqrt{2} \pi s} \left[\frac{4 u^2 \left(\bar{g}^{\ell}_A \pm \bar{g}^{\ell}_V\right)^2 C^\star_{LL,RR}  + 2 s^2\left((\bar{g}^{\ell}_V)^2 - (\bar{g}^{\ell}_A)^2\right)C^\star_{LR}}{t-\bar{M}_Z^2} + h.c \right].
  \eea
In the last two terms the $+/-$ in the expressions correspond to the left and right handed operators respectively.

\subsection{Partial widths extractions near and far from the $Z$ pole}\label{partialwidths}
Measured $e^+e^- \rightarrow \bar{f} f X$, $e^+e^- \rightarrow e^+e^- X$ inclusive processes at LEP are used to extract
values for the $Z$ decay partial widths assuming the SM. Here $X$ indicates the possible presence of photon or other final state emissions
that are not removed with hard isolation cuts. The strategy at LEP was to fit for the total width of the $Z$, ($\Gamma_Z$)
the $Z$ mass ($M_Z^2$), and a pole cross section ($\sigma_0$) as a function of center of mass energy scanning through the $Z$ pole.
Subsequently, ratios of cross sections are used to obtain partial decay widths for the $Z$. This approach is manifestly successful
as a hypothesis test of the SM. There is no statistically significant evidence that the SM breaks down in the EWPD program
when the SM is assumed.

When considering partial widths extracted from LEP data in the SM {\it at}  the $Z$ pole,
 $\sigma_{e^+e^- \rightarrow had}$ has the theoretical expression
\bea
\overline \sigma_h^0 =  12 \pi \, \frac{\overline \Gamma_{Z \rightarrow e \bar{e}} \overline \Gamma_{Z \rightarrow Had}}{|\overline \omega(M_Z^2)|^2},
\eea
 with $\overline \Gamma_{Z \rightarrow e \bar{e}}$, $\overline \Gamma_{Z \rightarrow Had}$ the decay in the SM.
 With the choice $\overline \omega(M_Z^2) = \bar{M}_Z \, \bar{\Gamma}_Z$,  and the partial width taking on SM values, this expression simplifies to the well known SM result.\footnote{
 Note that the SM result itself is neglecting contributions from the pure photon pole contribution, that are $\alpha_{ew}^2 \Gamma_Z^2/M_Z^2$ suppressed.}

\subsubsection{Partial widths in the SMEFT}\label{partial}
If one assumes that the SM does break down in the multi-TeV region and considers the general linear SMEFT,
the analysis path followed at LEP receives a number of corrections. These corrections include corrections of $\psi^4$ operators interfering
with the SM processes at tree level, and modifying the extracted $Z$ widths in the global data set.

The general correction to $\hat{\sigma}_h^0$ {\it near} the $Z$ pole ($s -M_Z^2 \equiv \Delta$) in the SMEFT is
\bea\label{hadronpole}
\frac{\delta \sigma_h^0}{\sigma_h^0} &\simeq& \frac{\delta \Gamma_{Z \rightarrow \ell \bar{\ell}}}{ \Gamma_{Z \rightarrow \ell \bar{\ell}}}
+  \frac{\delta \Gamma_{Z \rightarrow Had}}{ \Gamma_{Z \rightarrow Had}} - \frac{\delta \omega(M_Z^2)}{\omega(M_Z^2)}- \frac{\delta \omega^\star(M_Z^2)}{\omega^\star(M_Z^2)}.
\eea
where terms like : $\delta \sigma_{h, \psi^4}^0, \delta \sigma_{h, \gamma-Z}$, and $-2(\sigma_h^0)_{SM} \delta \omega/\omega$ are included into $\delta \sigma_h^0$.
For the near $Z$ pole hadronic cross section $\sigma(s)$ we have defined
\bea
\delta \sigma_{h, \psi^4}^0 = \left(2 \delta \sigma_{e^+ e^- \rightarrow u \bar{u}, \psi^4} + 3 \delta \sigma_{e^+ e^-  \rightarrow d \bar{d}, \psi^4} \right),
\eea
where
\bea
\delta \sigma_{e^+ e^- \rightarrow u \bar{u}, \psi^4} &=& \frac{N_c \hat{G}_F \hat{m}_Z^4}{6 \sqrt{2} \pi} \left[ \frac{\left(C_{\ell q}^{(1),\star} - C_{\ell q}^{(3),\star}\right)\left(G^{\ell}_V+G^{\ell}_A\right)\left(G^{u}_V + G^{u}_A\right)}{\Delta + i \, \omega(M_Z^2)}, \right. \\
&\,& \hspace{1.75cm} + \left. \frac{\left[(C_{eu}^\star+C_{\ell u}^\star) G^{\ell}_V + (C_{\ell u}^\star -C_{eu}^\star) G^{\ell}_A \right] \left(G^{u}_V - G^{u}_A\right)}{\Delta +i \, \omega(M_Z^2)} + h.c. \right], \nn
\delta \sigma_{e^+ e^- \rightarrow d \bar{d}, \psi^4} &=& \frac{N_c \hat{G}_F \hat{m}_Z^4}{6 \sqrt{2} \pi} \left[ \frac{\left(C_{\ell q}^{(1),\star} + C_{\ell q}^{(3),\star}\right)\left(G^{\ell}_V+G^{\ell}_A\right)\left(G^{d}_V + G^{d}_A\right)}{\Delta + i \, \omega(M_Z^2)},\right. \\
&\,& \hspace{1.75cm}
+ \left. \frac{\left[(C_{ed}^\star+C_{\ell d}^\star) G^{\ell}_V + (C_{\ell d}^\star -C_{ed}^\star) G^{\ell}_A \right] \left(G^{d}_V - G^{d}_A\right)}{\Delta+ i \, \omega(M_Z^2)} + h.c. \right].\nonumber
\eea
Here $G^{f}_{A/V}$ are the leading order predictions in the SM. Reintroducing flavour indicies is trivial in this case,
one finds $eeuu$ in all terms in the up quark case for example. Less trivial flavour indicies are present in the cases
with final state leptons and we find
\bea
\delta \sigma_{e^+_ie^-_i \rightarrow \nu_j \bar{\nu}_j,\psi^4} &=&  \frac{N_c \hat{G}_F \hat{m}_Z^4}{6 \sqrt{2} \pi} \left[ \frac{\left(G^{\ell}_V+G^{\ell}_A\right)}{\Delta + i \, \omega(M_Z^2)} \left(G^{\nu}_V + G^{\nu}_A\right)\left(C^\star_{\substack{\ell \ell \\ iijj}} + C^\star_{\substack{\ell \ell \\ ijji}} + C^\star_{\substack{\ell \ell \\ jiij}}\right), \right.  \\ &\,& \hspace{4.5cm} + \left. \frac{\left(G^{\ell}_V-G^{\ell}_A\right)}{\Delta + i \, \omega(M_Z^2)}\left(G^{\nu}_V + G^{\nu}_A\right) C^\star_{\substack{\ell e \\ iijj}}  + h.c
\right],\nn
\delta \sigma_{e^+ e^- \rightarrow e^+ e^-,\psi^4} &=& \frac{N_c \hat{G}_F \hat{m}_Z^4}{3 \pi \sqrt{2}} \left[2 \frac{\left(G_V^{\ell} + G_A^{\ell}\right)^2}{\Delta + i \omega(M_Z^2)} \left(C^\star_{\substack{\ell \ell \\ iijj}} + C^\star_{\substack{\ell \ell\\ ijji}} + C^\star_{\substack{\ell \ell \\ jiij}}\right), \right.  \\
&\,& \hspace{0.75cm}  + \left. 2 \frac{\left(G_V^{\ell} - G_A^{\ell}\right)^2}{\Delta + i \omega(M_Z^2)}\left(C^\star_{\substack{ee\\ iijj}}+C^\star_{\substack{ee\\ ijji}} +
C^\star_{\substack{ee\\ jiij}}\right) + \frac{(G_V^{\ell})^2-(G_A^{\ell})^2}{\Delta + i \omega(M_Z^2)} C^\star_{\substack{\ell e\\ iijj}} + h.c \right]. \nonumber
\eea
The correction $\delta \sigma_{h, \gamma-Z}$ is directly derivable from the previous results. As the effects of anomalous
$\gamma-Z$ interference terms have been studied in the literature to a larger degree, we do not discuss these corrections
in detail here.

Now consider the effect of the $\delta \sigma$ corrections due to $\psi^4$ operators in the combined global LEP data set, that includes $\sim 40 \, pb^{-1}$ of
data off the $Z$ peak, as well as $\sim 155 \, pb^{-1}$ of data at the $Z$ pole \cite{Z-Pole}. These $\psi^4$ corrections propagate
into the extracted partial widths and introduce theoretical errors when fits are performed in the SMEFT.

To illustrate these effects consider the expression for $\sigma_h^0$, where we can infer $ \Gamma_{Z \rightarrow Had}$,
assuming $\Gamma_{Z \rightarrow e^+ \, e^-}$ is a theoretical input. In this case
\bea
\delta \Gamma_{Z \rightarrow Had, \psi^4} &=& \frac{\hat{m}_Z^2 (\Gamma_Z^2)_{SM}}{12 \pi (\Gamma_{Z \rightarrow \ell \bar{\ell}})_{SM}} \delta \sigma_{h, \psi^4}^{(0)}.
\eea
Using $ \omega = M_Z \, \Gamma_Z$ one finds a correction to $\Gamma_{Z \rightarrow Had}$ of the form
\bea\label{inference1}
\delta \Gamma_{Z \rightarrow Had, \psi^4} &\simeq&  \left(\frac{(\Gamma_Z)_{SM} \, \hat{m}_Z}{\bar{v}_T^2} \right) \frac{\hat{m}_Z}{24 \, \pi^2 \, {\rm Br(Z \rightarrow e^+ \, e^-)}} \, \frac{\hat{m}_Z^2}{\bar{v}_T^2} \, C^{\psi^4} \, \frac{\bar{v}_T^2}{\Lambda^2}, \\
&\simeq& 0.01 \, {\rm GeV} \, C^{\psi^4} \, \frac{\bar{v}_T^2}{\Lambda^2}
\eea
Considering $\bar{v}_T^2/\Lambda^2 \sim 10^{-2}$ suppresses this correction to the order of the theoretical errors quoted for partial widths.
This indicates that the theoretical error introduced from such corrections in the SMEFT should not be completely neglected when precise bounds are of interest.

The leading effect of anomalous $Z$ couplings ($C^{\delta Z}$) introduce corrections to the partial widths that scale as
\bea
\delta \Gamma_{Z \rightarrow Had}&\simeq& \frac{\sqrt{2} \, \hat{G}_F \hat{m}_Z^3}{ \pi} C^{\delta Z} \, \frac{\bar{v}_T^2}{\Lambda^2}, \\
&\simeq& 3.98 \, {\rm GeV} \, C^{\delta Z}  \, \frac{\bar{v}_T^2}{\Lambda^2}
\eea
leading to a relative correction of the form
\bea
\frac{\delta \Gamma_{Z \rightarrow Had, \psi^4}}{\delta \Gamma_{Z \rightarrow Had}} \simeq  \left(\frac{(\Gamma_Z)_{SM} \, \hat{m}_Z}{\bar{v}_T^2} \right) \frac{1}{24  \, \pi  \rm Br(Z \rightarrow e^+ \, e^-) \, } \, \frac{C^{\psi^4}}{C^{\delta Z}}.
\eea

\subsubsection{Partial widths and ratios of cross sections}
The strategy employed at LEP is to extract partial widths in a global fit of EWPD pseudo-observables.
The global fit utilizes ratios of cross sections constructed out of the global data set, which includes off pole data.
The effect of $\delta \sigma$ corrections on this procedure can be characterized as introducing a correction of the form
\bea
\delta \frac{\sigma_{A\rightarrow B}}{\sigma_{C \rightarrow D}} \simeq \frac{ |\omega|^2}{\Gamma_C \, \Gamma_D} \delta \sigma^{\psi^4}_{AB} -
 \frac{|  \omega|^2 \, \Gamma_A \, \Gamma_B}{\Gamma_C^2 \, \Gamma_D^2} \delta \sigma^{\psi^4}_{CD}.
\eea
Here $\sigma_{AB}$ is an inclusive $A \rightarrow B$ cross section measurement which is constucted from data near the $Z$ pole.
Schematically $\Gamma_{A,B,C,D}$ are the partial decay widths inferred for the $Z$ from the ratios of cross sections,
and $C_{AB}$ stands for a $\psi^4$ operator that contributes. Using $ \omega = \Gamma_Z \, \hat{m}_Z$
and the scaling
\bea
 \delta \sigma^{\psi^4}_{AB} \simeq \frac{N_c \, \hat{m}_Z^2}{6 \pi \, \bar{v}_T^4} \, \frac{C_{AB} \, \bar{v}_T^2}{\Lambda^2},
\eea
one finds corrections to the extracted partial widths that are
\bea
\delta \frac{\sigma_{A\rightarrow B}}{\sigma_{C \rightarrow D}} &\simeq&
\frac{N_c \, \hat{m}_Z^4}{6 \, \pi  \, \bar{v}_T^4} \frac{1}{{\rm Br(Z \rightarrow C)}\, {\rm Br(Z \rightarrow D)}}
 \left[C_{AB} - \frac{\Gamma_A \, \Gamma_B}{\Gamma_C \, \Gamma_D} \, C_{CD} \right] \, \frac{\bar{v}_T^2}{\Lambda^2}, \nn
 &\simeq& 0.59 \,  \left( C_{AB} - C_{CD} \right) \,  \frac{\bar{v}_T^2}{\Lambda^2}
\eea
In the last step above we have taken all of the partial widths $\Gamma_{A,B,C,D}$ similar in size and the corresponding branching ratios
$\sim 10\%$ for illustrative purposes. Despite this dependence on $\psi^4$ operator Wilson coefficients, we emphasize the exact correction feeding into EWPD bounds
is very difficult to precisely quantify considering public data. We stress that this effect should not be over estimated.
Although the presence of unknown Wilson coefficients could contain hierarchies in some particular UV models,
it is unlikely that these corrections are significantly enhanced due to large Wilson coefficients.
The reason for this is the consistency checks at LEP included tests of anomalous $\gamma -Z$ interference terms.
As described in Ref \cite{Z-Pole} these consistency checks includes fitting for a nuisance parameter characterizing
an anomalous  $\gamma -Z$ interference term in off peak data at LEP. Further a joint analysis was
performed including lower energy ($\sqrt{s} = 58 \, {\rm GeV}$) data far off the $Z$ peak \cite{L3entry,Topaz,venus}. There is no evidence in these
results for large corrections to the $Z$ resonance shape. These consistency checks strongly imply
that in the case of the full SMEFT with anomalous $Z -\gamma$ interference and also $Z-\psi^4$ interference,
these terms are subdominant to leading order effects in possible anomalous $Z$ couplings to fermions.
The consistency checks reported by
LEP on anomalous $\gamma-Z$ interference do not place strong enough bounds on the anomalous interactions
to neglect these terms entirely in theoretical error estimates. See Refs. \cite{Kirsch:1994cf1,Kirsch:1994cf, Z-Pole,Isidori:1993ms} for further discussion.

We emphasize that our view is that this correction should be included as a theoretical error feeding into
a theoretical prediction in the SMEFT. The reasons for this are multifold. Firstly, for the SMEFT
it is reasonable to assume that
\bea
\frac{\bar{v}_T^2}{\Lambda^2} \sim \frac{\Gamma_Z \, M_Z}{\bar{v}_T^2}.
\eea
As such, neglected dimension eight operators would make directly fitting for
$\psi^4$ operators in the near $Z$ peak data suspect. Further perturbative corrections
to the higher dimensional operators are also comparable in size to corrections of this form.
We also emphasize that this correction is also further suppressed roughly by the fraction of off peak to $Z$ peak data
included in the global EWPD data set.
For these reasons, it is not advisable to fit for the $\psi^4$ operators
in near $Z$ pole data directly.

However, as all of these corrections
are present in the SMEFT, this makes introducing an extra theoretical error in fits
and adding it in quadrature with {\it the SM theoretical error} very well motivated.
In Section \ref{Numerics} we perform such a minimal EWPD fit.

\subsubsection{Near Z pole observables}\label{zpolesection}

In the SMEFT, at tree level, one has
\bea
 \bar{\Gamma} \left(Z \rightarrow f \bar{f} \right) &=& \frac{ \, \sqrt{2} \, \hat{G}_F \hat{m}_Z^3 \, N_c}{3 \pi} \left( |\bar{g}^{f}_V|^2 + |\bar{g}^{f}_A|^2 \right), \\
 \bar{\Gamma} \left(Z \rightarrow {\rm Had} \right) &=& 2 \, \bar{\Gamma} \left(Z \rightarrow u \bar{u} \right)+ 3  \, \bar{\Gamma} \left(Z \rightarrow d \bar{d} \right).
 \eea
With our chosen normalization of $\bar{g}_{V}^x = T_3/2 - Q^x \, \bar{s}_\theta^2, \bar{g}_A = T_3/2$ where $T_3 = 1/2$ for $u_i,\nu_i$ and $T_3 = -1/2$ for $d_i,\ell_i$
and $Q^x = \{-1,2/3,-1/3 \}$ for $x = \{\ell,u,d\}$.
The modification of the decay widths in the SMEFT compared to the situation in the SM introduces corrections of the form:
\bea
\delta \Gamma_{Z \rightarrow \ell \bar{\ell}}&=& \frac{\sqrt{2} \, \hat{G}_F \hat{m}_Z^3}{6 \pi} \, \left[ -\delta g^{\ell}_A + \left(-1 + 4 s_\that^2 \right) \delta g^{\ell}_V \right] + \delta�\Gamma_{Z \rightarrow \bar{\ell} \, \ell, \psi^4}, \\
\delta \Gamma_{Z \rightarrow \nu \bar{\nu}}&=& \frac{\sqrt{2} \, \hat{G}_F \hat{m}_Z^3}{6 \pi} \, \left[ \delta g^{\nu}_A +  \delta g^{\nu}_V \right] + \delta \Gamma_{Z \rightarrow \nu \bar{\nu},\psi^4} , \\
\delta \Gamma_{Z \rightarrow Had}&=& 2 \, \delta \Gamma_{Z \bar{u} u} + 3 \, \delta \Gamma_{Z \bar{d} d}, \\ &=& \frac{ \, \sqrt{2} \, \hat{G}_F \hat{m}_Z^3}{\pi} \left[ \delta g^{u}_A - \frac{1}{3} \left(- 3 + 8 s_{\that}^2 \right) \delta g^{u}_V - \frac{3}{2} \delta g ^{d}_A + \frac{1}{2}\left(- 3 + 4 s_{\that}^2 \right) \delta g^{d}_V  \right], \nonumber \\ &+& \delta \Gamma_{Z \rightarrow Had, \psi^4},
\eea
\bea
\delta \Gamma_{Z} &=& 3\delta \Gamma_{Z \rightarrow \ell \bar{\ell}} + 3 \delta \Gamma_{Z \rightarrow \nu \bar{\nu}} +\delta \Gamma_{had}, \\&=&  \frac{ \, \sqrt{2} \, \hat{G}_F \hat{m}_Z^3}{  2\, \pi} \left[ \delta g^{\nu}_A +  \delta g^{\nu}_V -\delta g^{\ell}_A + \left(-1 + 4 s_\that^2 \right) \delta g^{\ell}_V, \right. \nonumber \\ &\, & \hspace{2.2cm} \left. + 2 \delta g^{u}_A - \frac{2}{3} \left(- 3 + 8 s_{\that}^2 \right) \delta g^{u}_V - 3 \delta g ^{d}_A + \left(- 3 + 4 s_{\that}^2 \right) \delta g^{d}_V \right], \nonumber \\ &+& \delta \Gamma_{Z \rightarrow Had, \psi^4} + 3 \delta \Gamma_{Z \rightarrow \ell \bar{\ell}, \psi^4} + 3 \delta \Gamma_{Z \rightarrow \nu \bar{\nu},\psi^4}.
\eea
So that: $ \bar{\Gamma} \left(Z \rightarrow f \bar{f} \right)= \Gamma_{Z \rightarrow f \bar{f}} + \delta \Gamma_{Z \rightarrow f \bar{f}}$ for all $f$ and the same kind of relation holds for  $\bar{\Gamma}_{Z}$.
The shift of the ratios of decay rates defined in the SM as $R^0_{f}=\frac{\Gamma_{had}}{\Gamma_{Z \rightarrow \bar{f} f}}$ where $f$ can be a charged lepton $\ell$ or a neutrino
follows from
\bea
\delta R^0_{f}=\frac{1}{(\Gamma(Z \rightarrow f \bar{f})^2)_{SM}} \left[ \delta \Gamma_{Z \rightarrow Had} (\Gamma(Z \rightarrow f \bar{f}))_{SM} - \delta \Gamma_{Z \rightarrow f \bar{f}}  (\Gamma \left(Z \rightarrow {\rm Had})_{SM} \right)\right],
\eea
and we can then write that $\bar{R}^0_{f}= R^0_{f} + \delta R^0_{f}$. For an identified quark the inverse ratio is used.
\subsubsection{Forward Backward asymmetry}
The forward backward asymmetry for 2-2 scattering is defined as
\bea
A_{FB} = \frac{\sigma_F - \sigma_B}{\sigma_F + \sigma_B}.
\eea
Here $\sigma_F$ is defined by $\theta \in \left[0,\pi/2\right]$ and $\sigma_B$ is defined by $\theta \in \left[\pi/2, \pi \right]$ with $\theta$ defined as
in Section \ref{differential}. In the SM, it can be shown that the forward backward asymmetry for leptons is just
\bea
A_{FB}^{0, f}= \frac{3}{4} A_{e} A_{f}, \quad A_{e}= 2 \frac{g^{\ell}_V g^{\ell}_A}{ (g^{\ell}_V)^2 + (g^{\ell}_A)^2}, \quad A_{f}= 2 \frac{g^{f}_V g^{f}_A}{ (g^{f}_V)^2 + (g^{f}_A)^2}.
\eea
As we move to the SMEFT, the $Z$ couplings receive corrections bringing corrections to $A_{FB}^{0,f}$. $A_{FB}^{0,f}$ also receive corrections from  $\omega$ redefinition in general, and from $\psi^4$ operators. All of these corrections can be derived from Eqn \ref{generaldifferential}, but we note the following simplified expressions.
In the SMEFT $\bar{A}_{f}$ can be written as
\bea
\bar{A}_f = \frac{2 \bar{r}_f}{1 + \bar{r}_f^2},
\eea
where $\bar{r}_f = \frac{\bar{g}^{f}_V}{\bar{g}^{f}_A}$. The redefinition of the $Z$ coupling then leads to a shift of $\bar{A}_f$ such that $\bar{A}_f = (A_{f})_{SM} \left( 1 + \frac{\delta A_{f}}{(A_{f})_{SM}}\right)$ where
\bea
\frac{\delta A_{f}}{(A_{f})_{SM}} = \delta r_f \left( 1 - \frac{2 (r_{f}^2)_{SM}}{1+ (r_{f}^2)_{SM}}\right).
\eea
Here $\delta r_f$ is defined by $r_f = (r_{f})_{SM} \left( 1 + \delta r_{f} \right)$ with $\delta r_{f} =  \delta g^{f}_V/ G^{f}_V - \delta g^{f}_A/ G^{f}_A$. We again use : $(...)_{SM}$ for leading order SM predictions and $G^{f}_{A,V}$ for leading order SM predictions for the couplings.
Then the corrections to $A_{FB}^{0,f}$ from the shifts in the effective couplings are
\bea
\delta A_{FB}^{0,f} =\frac{3}{4} \left[ \delta A_{\ell} \,  (A_{f})_{SM} + (A_{\ell})_{SM} \, \delta A_{f}\right].
\eea
The corrections due to $\psi^4$ operators $\delta (A_{FB}^{0,f})_{\psi^4}$ and the redefinition of $\omega$ can be extracted from:
\bea
\frac{3}{4} (A_{\ell} A_{f})_{SM} \left( \frac{\delta \left( \sigma_F - \sigma_B \right)}{(\sigma_{F} - \sigma_{B})_{SM}} - \frac{\delta \left( \sigma_F + \sigma_B \right)}{(\sigma_{F} + \sigma_{B})_{SM}}  \right),
\eea
where the contributions $\delta \left( \sigma_F - \sigma_B \right)$, $\delta \left( \sigma_F + \sigma_B \right)$ that depend on $\psi^4$ operators, are derived directly from Eqn \ref{generaldifferential}. As the forward backward asymmetry measurements are direct cross section measurements, the scaling of Section \ref{scaling} holds and these
$\psi^4$ corrections can be neglected for near $Z$ pole analyses. Far off the $Z$ pole, these corrections cannot be neglected. In particular, in interpreting reported $A_{FB}$ measurements
reported with LEPII data, these corrections are not suppressed compared to the effects of anomalous $Z$ couplings.

\section{Numerics}\label{Numerics}

In this Section we perform some minimal EWPD fits.
The results presented here are not intended to be a global analysis of all possible data. Our purpose is to make clear
a number of challenges present in such fit efforts in the SMEFT that have not been discussed in the literature,
including the neglect of the effects we have discussed in some detail in Section \ref{partial}.
We then suggest an approach to circumvent a number of these challenges in Section \ref{EWtoLHC}.
The Wilson coefficients (naively) present in the set of observables we examine are
\bea
C_{fit} = \frac{\bar{v}_T^2}{\Lambda^2}\{C_{\substack{H q \\ pr}}^{(1)}, C_{\substack{H q \\ pr}}^{(3)}, C_{\substack{H u \\ pr}}, C_{\substack{H d \\ pr}} , C_{\substack{H \ell \\ pr}}^{(1)}, C_{\substack{H \ell \\ pr}}^{(3)}, C_{\substack{H e \\pr}}, C_{\substack{ll}}, C_{HD} ,  C_{HWB}\}.
\eea
In the $\rm U(3)^5$ limit, there are ten parameters in the set of nine measurements given in Table  \ref{EWtable}. Field redefinitions to remove an operator
do not effect physical measurements, and cannot lead to a more constrained field theory. We do not attempt to remove parameters
by field redefinitions to match the number of parameters and measurements\footnote{Such a choice is meaningless in the SMEFT, which has an infinite number of parameters in general.}, but simply construct the $\chi^2$ directly.

We construct a $\chi^2$ for a EWPD fit in the following way.
We define a matrix $\mathcal{C}$ as the covariance matrix of the
observables, the experimental values of which are obtained from Ref. \cite{Z-Pole} . $\Delta \, \theta_i$ as a vector of the difference
in the observed and predicted value of an observable, as a function of the unknown Wilson coefficients. The $\chi^2$ is then given by
\bea
\chi^2_{EW} = (\Delta \theta_i)^T  \, (\mathcal{C}^{-1})_{i,j} \, (\Delta \theta_j).
\eea
The minimum $\chi^2_{EW,min}$ is determined, and the $65 \%, 90 \%$ and
$99 \%$ best fit confidence level regions ($\Delta \chi^2_{EW}$) are defined by the cumulative distribution
function for a multi-parameter fit. The confidence level regions are then
given by  $\chi^2_{EW} = \chi^2_{EW, min} + \Delta \chi_{EW}^2$.

For theoretical predictions in the SM, we use the results supplied by the updated 2013 PDG \cite{Agashe:2014kda}
and Ref.\cite{Freitas:2014hra}. We do not use as SM predictions the results of a fit to EWPD observables.
Minimized fit results of this form for the SM (with a number of SM parameters floated as in \cite{Baak:2014ora}) is a valid procedure
for hypothesis testing the SM. When considering a fit in the SMEFT, using such fit values as the SM theoretical predictions
is only valid if the corrections due to unknown Wilson coefficients enter into the combined $\chi^2$ in a manner that does not depend on the SM
parameters fit to themselves. This is an unvalidated assumption in the SMEFT, and as such we use the SM predictions supplied by
\cite{Agashe:2014kda, Freitas:2014hra}.

\begin{center}
\begin{table}
\centering
\tabcolsep 8pt
\begin{tabular}{|c|c|c|c|c|}
\hline
Observable & Experimental Value & Ref. & SM Theoretical Value & Ref.  \\ \hline
$\hat{m}_Z $[GeV] & $91.1875 \pm 0.0021$ & \cite{Z-Pole} &-&-\\
$M_W $[GeV] & $80.385 \pm 0.015 $ & \cite{Group:2012gb} &$80.365 \pm 0.004$& \cite{Awramik:2003rn}\\
$\Gamma_{Z}$[GeV] &$2.4952 \pm 0.0023 $&\cite{Z-Pole} &$2.4942 \pm 0.0005$& \cite{Freitas:2014hra}\\
$R_{\ell}^{0}$ & $20.767 \pm 0.025$ &\cite{Z-Pole} &$20.751 \pm 0.005$& \cite{Freitas:2014hra}\\
$R_{c}^{0}$ & $0.1721 \pm 0.0030$ &\cite{Z-Pole} &$0.17223 \pm 0.00005$& \cite{Freitas:2014hra}\\
$R_{b}^{0}$ & $0.21629 \pm 0.00066$ &\cite{Z-Pole} &$0.21580 \pm 0.00015$&\cite{Freitas:2014hra}\\
$\sigma_h^{0}$ [nb]& $41.540 \pm 0.037$ & \cite{Z-Pole}&$41.488 \pm 0.006$& \cite{Freitas:2014hra}\\
$A_{FB}^{\ell}$ & $0.0171 \pm 0.0010$ & \cite{Z-Pole}&$0.01616 \pm 0.00008 $&\cite{Agashe:2014kda} \\
$A_{FB}^{c}$ & $0.0707 \pm 0.0035$ & \cite{Z-Pole}&$0.0735 \pm 0.0002$& \cite{Agashe:2014kda} \\
$A_{FB}^{b}$ & $0.0992 \pm 0.0016$ & \cite{Z-Pole} &$0.1029 \pm 0.0003$& \cite{Agashe:2014kda} \\
\hline\end{tabular}
\caption{Experimental and theoretical values of the observables used in the illustrative fits.\label{EWtable}}
\end{table}
\end{center}

\subsection{Prior dependence}
We find that obtaining a global minimum, and hence a detailed fit space for the unknown Wilson coefficients ($C_{fit}$) is numerically unstable
and strongly depends on the seed imposed in the search and the priors used.\footnote{A further very basic problem for consistency
in the SMEFT is for any minima to be obtained, cross terms of order $v^4/\Lambda^4$ need to be included in the $\chi^2_{EW}$.This is while terms
from dimension eight operators are neglected, that can appear. As we argue, including an extra theoretical error for these neglected terms is more consistent
than effectively treating the SMEFT as exactly $\mathcal{L}_{SM} + \lsix$.}
This is not surprising as the number of unknown Wilson coefficients
present in the SMEFT is large.

For example, a set of reasonable prior conditions to impose is that the power counting expansion of the theory is under control,
and that each individual observable falls within $N \sigma$ of each measurement, so that
\bea\label{conditions}
C_{fit}  <  0.1, \quad \quad \hat{\theta}_i - \theta_i(C_{fit}^{min})   < N \, \delta \theta_i
\eea
with $\delta \theta_i$ the total combined error on an observable $\theta_i$. The value of $N$ chosen in these conditions
dictates the specific global minimum found in the $\chi^2$ minimization. In particular the presence of the $A_{FB}^b$ anomaly that deviates at the
$\sim 2.5 \, \sigma$ level from the SM predictions indicates that $N > 2.5$ as a minimization condition is reasonable to not
bias the global minimum in favour of non-vanishing $C_{fit}^{min}$. Choosing $N = 2.8$,  and seeding a minimization with $C_{fit}^{min} =0$, we find
\bea
C_{fit}^{min} = \{0.5, -0.3, 8.9, -31,  1.1, 3.3, 0.6, -1.4, 1.1, -1.9\}  \times 10^{-3}.
\eea
It is interesting to note that with this procedure some of the least constrained entries in $C_{fit}^{min}$
corresponds to operators that lead  to vertex corrections of the $Z$ boson to fermions.

However, we stress the arbitrariness of the conditions imposed
to obtain this minima and that it does not hold any particular physical significance.
For example, another reasonable prior condition can be constructed based on noting that
one can group the $C_i$ into subgroups that strongly mix under RG evolution
(see Ref.\cite{Jenkins:2013zja,Jenkins:2013wua,Alonso:2013hga,Alonso:2014zka} for the relevant RGE results). Such Wilson coefficients will
tend to flow together in value under RG evolution. This can motivate grouping the operators into classes of the form
\bea
C_q &=& \{C_{\substack{H q \\ pr}}^{(1)}, C_{\substack{H q \\ pr}}^{(3)}, C_{\substack{H u \\ pr}} C_{\substack{H d \\ pr}} \},  \quad \quad
C_\ell = \{C_{\substack{H \ell \\ pr}}^{(1)}, C_{\substack{H \ell \\ pr}}^{(3)}, C_{\substack{H e \\pr}} \}.
\eea
Then imposing the conditions in Eqn.\ref{conditions} gives a minimum with these grouped Wilson coefficients $\mathcal{O}(10^{-5})$
and $C_{HWB} \sim \mathcal{O}(10^{-4})$.
The individual minima, with two different prior conditions significantly differ. The allowed fit space is also highly prior dependent.
As such, fitting for a best fit value of an individual Wilson coefficient only allows weak conclusions to be drawn.
Marginalizing over all unknown Wilson coefficients in EWPD, or a subset of measurements, introduces further prior
dependence. In reasonable UV scenarios, the unknown Wilson coefficients are expected to be extremely
highly correlated. Using a prior condition to remove cases where correlations between Wilson coefficients
allow larger values in the unknown parameters is poorly motivated for this reason. Unfortunately, at the same time, the particular correlations in $\lsix$ for all possible
UV models is unknown.

\subsection{Theoretical Errors in the SMEFT}
The fit space of allowed Wilson coefficients is strongly prior dependent. In particular, we find that the condition that
$\hat{\theta}_i - \theta_i(C_{fit}^{min})   < N \, \delta \theta_i$
implicitly or explicitly being imposed strongly dictates the allowed Wilson coefficient space. For this reason, a precise specification of the
theoretical error when fitting in the SMEFT is critical.

It is essential to distinguish between the cases
of fitting to EWPD as a hypothesis test of the SM itself, and fitting to EWPD assuming the SMEFT as a theoretical framework.
When using EWPD to  hypothesis test the SM, theoretical errors for unknown higher order SM corrections
are specified and included in a fit.
Adding higher dimensional operators to a fit of this form can also be interpreted as a (less efficient) hypothesis test of the SM,
if no extra theoretical error is added. For sample fits of this form see Ref \cite{Pomarol:2013zra,Falkowski:2014tna}.

Conversely if the theory assumed in an EWPD fit is the SMEFT, the theoretical error differs from the SM. Extra theoretical errors
should be added in quadrature to the SM theoretical errors when bounds on Wilson coefficients are extracted.
This is particularly required if constraints on Wilson coefficients are to be used at LHC as a
test of the linear SMEFT formalism itself.\footnote{An important example of studies of this form is the constraints from EWPD projected onto the $h \rightarrow V \, F$ spectra, which are de-correlated in
the case of the nonlinear EFT from LEP measurements\cite{Isidori:2013cga,Brivio:2013pma}.} The SMEFT is subject to substantial theoretical errors of this form.
There are three major sources of error:
\begin{itemize}
\item
The full dependence of EWPD $2 \rightarrow 2$ scattering processes in the SMEFT is now systematically
characterized to leading order in $1/\Lambda^2$, with the results in
Section \ref{obs}. We have shown this  introduces dependence on higher dimensional operators
suppressed by $\Gamma_Z \, M_Z/\bar{v}_T^2$ compared to the leading order effect suppressed
by  $\bar{v}_T^2/\Lambda^2$ in extracted partial widths. This error does not effect all
processes equally in EWPD, which distorts $\chi^2_{SMEFT }$ compared to $\chi^2_{SM}$.
Similar comments hold for the effect of $Z -\gamma$ interference in near pole $Z$ data.

\item
Neglected perturbative corrections in the SMEFT.
Although the full RGE results of the SMEFT dimension six operators
are now known, perturbative corrections in EWPD for higher dimensional operators are generally neglected.
The neglected perturbative corrections are of the order
\bea
\frac{\bar{v}_T^2}{\Lambda^2} \frac{\bar{g}^2}{16 \, \pi^2} \sim \mathcal{O}(10^{-3}) \frac{\bar{v}_T^2}{\Lambda^2}
\eea
for $\bar{g}_1, \bar{g}_2$ corrections in the SMEFT. The corrections are an order of magnitude larger
for QCD effects. Perturbative corrections to the Wilson coefficients in $C_{fit}$ can be absorbed into the unknown Wilson coefficient.
However, perturbative corrections of this form also introduce a dependence on a large number of higher dimensional operators that
are not in the set $C_{fit}$. These corrections should be treated as a theoretical error when extracting bounds
on $C_{fit}$ to use in other measurements.

\item
Neglect of dimension eight operators introduces theoretical errors of the order
\bea
\frac{\bar{v}_T^4}{\Lambda^4} \sim \mathcal{O}(10^{-2}) \frac{\bar{v}_T^2}{\Lambda^2}
\eea
for $\Lambda \sim {\rm TeV}$. These corrections cannot be simply absorbed into a set of effective
$C_{fit}$ parameters if the bounds obtained in EWPD are to be used in another process.
\end{itemize}

As an illustrative example of the importance of including the theoretical error of the SMEFT consistently, consider the case
of near pole corrections due to $\psi^4$ operators. These corrections modify extracted partial widths. Including a universal extra theoretical error $\delta_E$
in the partial widths one finds
\bea
\frac{\delta \chi_{EW}^2}{\delta_E^2} + 10^{6}&=& 10^{8} \, \frac{\bar{v}_T^2}{\Lambda^2} \left(
-6.5 \, C_{HWB} + 4.5\,C_{Hq}^{(3)} -3.4 \, C_{HD} + 4.5 \,C_{H\ell}^{(1)}, \right. \nn
&\,&\hspace{1.5cm} \left. + 6.8 \,C_{\ell \ell} - 1.3 \, C_{Hd} - 1.7 \, C_{Hu} -1.1 C_{He}, \right. \nn
&\,&\hspace{1.5cm} \left. 1.4 \,C_{Hq}^{(1)} -7.9 \, C_{H\ell}^{(3)}+ \cdots \right),
\eea
even though $\delta_E \sim 10^{-3}$ corrections of this form significantly modify any extracted constraints.\
This is easily seen by direct inspection of the leading terms in the $\chi^2$, which are
\bea\label{minimal}
\chi^2_{EW} - 11.1 &=&  10^{3} \frac{\bar{v}_T^2}{\Lambda^2} \left(5.4 \, C_{HWB} - 1.2  \,C_{Hq}^{(3)} + 4.7 \, C_{HD}
- 5.6 \,C_{H\ell}^{(1)},  \right. \\ &\,&\hspace{1.5cm} \left.- 2.7  \,C_{\ell \ell} + 0.61   \, C_{Hd} -0.2 \, C_{Hu} + 6.6 \, C_{He} + 0.9 \,C_{Hq}^{(1)}
 + 4.0 \, C_{H\ell}^{(3)}+ \cdots \right). \nonumber
\eea
These corrections change the vector of $C_{fit}$ that is constrained by EWPD.
For this reason it is
important to carefully account for theoretical error when fitting in the SMEFT to explore patterns of allowed deviations.
$\delta_E$ is not a universal shift in the SMEFT in a full analysis, but depends on different $\psi^4$ operators.
This can change the the vector of $C_{fit}$ that is constrained by EWPD in an even more dramatic fashion.
For this reason, it is important to also incorporate correlated constraints on $\psi^4$ operators in the SMEFT in global fits.

All of these corrections introduce theoretical errors in the SMEFT and can be enhanced
by unknown order one Wilson coefficients.  For all of these reasons leading order bounds on $C_i \bar{v}_T^2/\Lambda^2$ that exceed the  $\mathcal{O}(10^{-2})$ level are challenging to interpret as consistent constraints on parameters in $\lsix$.

\subsection{Relating EW $\chi^2$ constraints to LHC processes}\label{EWtoLHC}
Due to the challenges we have discussed on the usual procedure to fit to parameters in $\lsix$ it is of interest to
have a viable alternative to project EW precision constraints onto the LHC program. In this Section, we argue that such an alternative is supplied {\it by directly running
the $\chi^2_{EW}$ constraint to LHC energies} and then imposing it on a related processes in the linear SMEFT.

Naively one might argue that the running of the Wilson coefficients can be neglected as
such perturbative corrections are on unknown parameters. However,
when running the -- $\chi^2_{EW}$ function ---
this argument
fails by direct inspection of the $\chi^2_{EW}$ dependence on the Wilson coefficients. As can be seen in Eqn \ref{minimal}
the numerical factors that multiply the unknown Wilson coefficients are hierarchical and differ by orders of magnitude.
As such interpreting an EWPD constraint as
\bea
\chi^2_{EW}(m_Z) \equiv \chi^2_{EW}(m_h)
\eea
for the sake of the constrained Wilson coefficients at the scale $m_h$ is inaccurate and actually constrains the wrong set of parameters.

Alternatively, consider running $\chi^2_{EW}$ as a constraint vector in the Wilson coefficient space to LHC energies. This shows that the constraint vector
then depends on {\it different} unknown Wilson coefficients, with a comparable numerical pre-factor to the coefficients present in $C_{fit}$ at the scale $m_Z$.
In other words, the directions in Wilson coefficients space
constrained at LEP are rotated evolving up to LHC energies and this rotation does not leave the constraint vector on the same Wilson coefficient
Hypersurface.
The large hierarchies in the numerical coefficients that define the $\chi^2$ enhance the effects of RGE running even from $m_Z$ to $m_h$
when considering constraints on Wilson coefficients, and as a result this is not a negligible effect.

\begin{figure}
\hspace{3.5cm}
\begin{tikzpicture}
\draw[decorate,decoration=snake] (0,0) -- (1.5,0) ;
\filldraw (1.4,-0.1) rectangle (1.6,0.1);
\draw [->](1.5,0) -- (2.1,0.45);
\draw (2.1,0.45) -- (2.5,0.75);
\draw  (1.5,0) -- (2,-0.375);
\draw [<-](2,-0.375) -- (2.5,-0.75);
\node [above] at (0.5,0) {$Z$};
\end{tikzpicture}
\hspace{1cm}
\begin{tikzpicture}
\draw[decorate,decoration=snake] (0,0) -- (1.5,0) ;
\draw (2,0) circle [radius=0.5];;
\filldraw (2.4,-0.1) rectangle (2.6,0.1);
\draw [->](2.5,0) -- (3.0,0.375);
\draw (3.0,0.375) -- (3.5,0.75);
\draw [->](2.5,0) -- (3.0,-0.375);
\draw (3.0,-0.375) -- (3.5,-0.75);
\node [above]at (0.5,0) {$Z$};
\draw [->] (1.99,0.5) -- (2,0.5);
\draw [<-] (1.99,-0.5) -- (2,-0.5);
\end{tikzpicture}
\caption{\label{fig:2}
Mixing of a $\psi^4$ operator into an effective $Z$ coupling to fermions.}
\end{figure}
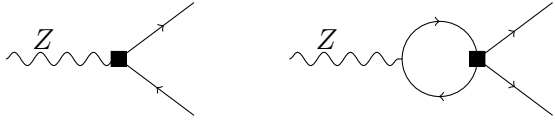

For example, consider the running of $C_{HWB}$ due to $y_t$. Using the results in Ref \cite{Jenkins:2013wua}
\bea
\mu \frac{d \, C_{HWB}}{d \, \mu} = -\frac{2 \, g_1 \, N_c (\hyp_q + \hyp_u) \, y_t}{16 \, \pi^2} {\rm Re} (C_{\substack{uW \\ 33}}) + \cdots
\eea
with $ \hyp_q,\hyp_u$ the $q$ and $u$ hypercharges. This introduces dependence on $ {\rm Re} (C_{\substack{uW \\ 33}})$
of the form
\bea
\Delta \chi_{EW}^2(m_h) \sim - 10^{3} \frac{\bar{v}_T^2}{\Lambda^2} \,  \left(0.1 \, {\rm Re} (C_{\substack{uW \\ 33}}) \cdots \right)
\eea
into $\chi^2(m_h)$.
Such dependence is comparable to the dependence of a number of the remaining Wilson coefficients in $C_{fit}$ in $\chi^2_{EW}(m_Z)$.
As a further illustrative example, consider running an effective $Z$ coupling to fermions, such as $\delta g^{x}_{V,A}$.
This interaction receives further four quark operator corrections at the one loop level as shown in Fig \ref{fig:2}.
Extracting a related result for the leading log running directly from Ref \cite{Alonso:2013hga}
\begin{align*}
\mu \, \frac{d}{d \, \mu} C^{(1)}_{\substack{H l \\ rs}} &= \frac{1}{48 \, \pi^2} g_1^2 \hyp_H \left(
\hyp_H   C_{\substack{Hl \\ rs}}^{(1)} + N_c \hyp_d  C_{ \substack {ld \\ r s w w } }
+ \hyp_e  C_{ \substack {le \\ r s w w } }
+2 \hyp_l C_{ \substack {ll \\ r s w w } }
+ \hyp_l C_{ \substack {ll \\ r w w s } } \right.,
\nn
& \left. \hspace{2.5cm}
+ \hyp_l C_{ \substack {ll \\ w s r w } }
+2 \hyp_l C_{ \substack {ll \\ w w r s } }
+2 N_c \hyp_q C^{(1)}_{ \substack {lq \\ r s w w } }
+ N_c  \hyp_u C_{ \substack {lu \\ r s w w } } \right).
\end{align*}
In the $\rm U(3)^5$ limit the operators in $\delta g^{\ell}_{V,A}$ mix with a total of ten four fermi operators
\cite{Jenkins:2013wua,Alonso:2013hga}.
Taking into account such effects in running  $\chi^2_{EW}(M_Z)$ to $\chi^2_{EW}(m_h)$ makes clear it is essential to
perform a global analysis, including constraints on $\psi^4$ operators if one is interested in projecting EW
constraints to LHC processes.\footnote{For some studies of constraints on $\psi^4$ operators see Ref \cite{GonzalezGarcia:1998ay}. It is interesting to
note that operators of the form considered in detail for off-pole $\psi^4$ corrections do not induce anomalous magnetic moments
of leptons due to their chirality.} At the scale $m_h$ a prior dependent minimization, and possibly a marginalization of the Wilson coefficients subject to $\chi^2_{EW}$
is still required. However this approach allows multiple measurements at different scales to be evolved and
directly combined into a global constraint $\chi^2$. This occurs before one global minimization and marginalization is preformed, minimizing the prior dependence.

\section{Conclusions}\label{conclusions}
The SMEFT has many unknown parameters that have been probed in particular combinations at many different energy scales.
It is important to incorporate the bounds from EWPD when searching for deviations from the SM in the LHC program.
However, at the same time it is important to consistently incorporate constraints from EWPD, and to not artificially
increase the strength of bounds in an inconsistent analysis. For this reason,
it is essential to include theoretical error for the SMEFT itself in fitting to $\lsix$
to explore patterns of allowed deviations. This is the case if the assumption is that the theory being constrained is the SMEFT. We have shown that constraints in the SMEFT from EWPD are subject to theoretical uncertainties
that have been neglected in previous analyses. Our general results for LEPI and LEPII scattering cross sections enable global EWPD constraint efforts in the SMEFT
to advance further, and help characterize this theoretical error.

We have discussed some challenges present in naively utilizing EWPD fit constraints
in the SMEFT.
We have also argued for running a constraint $\chi^2_{EW}$ using RG evolution and directly applying it to related processes occurring at different energy scales.
This is preferred over minimizing and marginalizing to fit for individual Wilson coefficients at the scale $m_Z$, when ignoring perturbative corrections
in the SMEFT.
The requirement to construct a consistent global constraint picture of the linear SMEFT remains urgent as the LHC physics program advances.

\section*{Acknowledgements}

We thank Alberto Guffanti for many interesting conversations and Gfitter intervention.
We also thank  Aneesh Manohar and Gino Isidori for comments on the manuscript.
We thank Mikkel Bj\o rn for pointing out an error in Section 3, and Ilaria Brivio for spotting a typo in Section 4.
M.T. acknowledges generous support by the Villum Fonden and partial support by the Danish National Research Foundation (DNRF91).
The project leading to this application has received funding from the European Union's Horizon 2020 research
and innovation programme under the Marie Sklodowska-Curie grant agreement No 660876, HIGGS-BSM-EFT.

\appendix
\section{Operators and Notation Used}
The operators that have contributed to corrections in the SMEFT, that were not explicitly defined in the
body of the paper are
\begin{align}
Q_H &= (H^\dag H)^3, \quad \quad  &Q_{H D} &= \left(H^\dag D_\mu H\right)^* \left(H^\dag D_\mu H\right), \\
Q_{H W} &= H^\dag H\, W^I_{\mu\nu} W^{I\mu\nu},  \quad \quad &Q_{H WB} &= H^\dag \tau^I H\, W^I_{\mu\nu} B^{\mu\nu}, \\
Q_{\substack{H l \\pr}}^{(1)} &= (H^\dag i\overleftrightarrow{D}_\mu H)(\bar l_p \gamma^\mu l_r),
\quad \quad &Q_{\substack{H l \\pr}}^{(3)} &=(H^\dag i\overleftrightarrow{D}^I_\mu H)(\bar l_p \tau^I \gamma^\mu l_r),\\
Q_{\substack{H e \\pr}} &= (H^\dag i\overleftrightarrow{D}_\mu H)(\bar e_p \gamma^\mu e_r),
\quad \quad &
Q_{\substack{H q \\pr}}^{(1)} &= (H^\dag i\overleftrightarrow{D}_\mu H)(\bar q_p \gamma^\mu q_r), \\
Q_{\substack{H q \\pr}}^{(3)} &= (H^\dag i\overleftrightarrow{D}^I_\mu H)(\bar q_p \tau^I \gamma^\mu q_r),
\quad \quad & Q_{\substack{H u \\pr}} &= H^\dag i\overleftrightarrow{D}_\mu H)(\bar u_p \gamma^\mu u_r), \\
Q_{\substack{H d \\pr}} &=(H^\dag i\overleftrightarrow{D}_\mu H)(\bar d_p \gamma^\mu d_r), \quad \quad
& Q_{\substack{H u d \\pr}}  &= i(\widetilde H ^\dag D_\mu H)(\bar u_p \gamma^\mu d_r), \\
Q_{\substack{uW \\pr}} &=(\bar q_p \sigma^{\mu\nu} u_r) \tau^I \widetilde H \, W_{\mu\nu}^I.
\end{align}

Here we have used the derivative notation
\begin{align}
H^\dagger \, i\overleftrightarrow D_\beta H &= i H^\dagger (D_\beta H) - i (D_\beta H)^\dagger H, \\
H^\dagger \, i\overleftrightarrow D_\beta^I H &= i H^\dagger \tau^I (D_\beta H) - i (D_\beta H)^\dagger \tau^I H.
\end{align}

The Lagrangian we use is given by $\mathcal{L}=\mathcal{L}_{\rm SM}+ \mathcal{L}^{(5)}+\lsix + \cdots$, where $\lsix = \Sigma_i C_i \, Q_i$. To establish notation,
we note  $H$ is an $\rm SU(2)$ scalar doublet with hypercharge $\hyp_H=1/2$. The Higgs boson mass is given as $m_H^2=2\lambda \bar{v}_T^2$, with $\bar{v}_T \sim 246$~GeV. The covariant derivative is $D_\mu = \partial_\mu + i g_3 T^A A^A_\mu + i g_2  t^I W^I_\mu + i g_1 \hyp B_\mu$. Here $T^A$ are $\rm SU(3)$ generators,  $t^I=\tau^I/2$ are $SU(2)$, and $\hyp$ is the $\rm U(1)$ Hypercharge generator.  $\widetilde H$ is defined by $H_j = \epsilon_{jk} H^{\dagger\, k}$
where the $\rm SU(2)$ invariant tensor $\epsilon_{jk}$ is defined by $\epsilon_{12}=1$ and $\epsilon_{jk}=-\epsilon_{kj}$, $j,k=1,2$.  Fermion fields $q$ and $l$ are left-handed fields, and $u$, $d$ and $e$ are right-handed fields. We use $p,r,s,t$ for flavor indices. The effective mixing angles are defined as
\begin{align}
\sin \tc &= \frac{\gcb}{\sqrt{\gcb^2+\gcw^2}}\left[1 + \frac{\bar{v}_T^2}{2}  \,   \, \frac{\gcw}{\gcb}\ \frac{\gcw^2-\gcb^2}{\gcw^2+\gcb^2} C_{HWB} \right],  \\
\cos \tc &= \frac{\gcw}{\sqrt{\gcb^2+\gcw^2}}\left[1 - \frac{\bar{v}_T^2}{2}  \,   \, \frac{\gcb}{\gcw}\ \frac{\gcw^2-\gcb^2}{\gcw^2+\gcb^2} C_{HWB} \right].
\end{align}
The formalism of the paper for the SMEFT, and some results used in Section \ref{inputs} descend from Refs \cite{Jenkins:2013zja,Alonso:2013hga}.

\bibliographystyle{JHEP}
\bibliography{RGcopy}

\providecommand{\href}[2]{#2}\begingroup\raggedright\begin{thebibliography}{10}

\bibitem{Grinstein:1991cd}
B.~Grinstein and M.~B. Wise, {\it {Operator analysis for precision electroweak
  physics}},  {\em Phys.Lett.} {\bf B265} (1991) 326--334.

\bibitem{Han:2004az}
Z.~Han and W.~Skiba, {\it {Effective theory analysis of precision electroweak
  data}},  {\em Phys.Rev.} {\bf D71} (2005) 075009,
  [\href{http://xxx.lanl.gov/abs/hep-ph/0412166}{{\tt hep-ph/0412166}}].

\bibitem{Ciuchini:2013pca}
M.~Ciuchini, E.~Franco, S.~Mishima, and L.~Silvestrini, {\it {Electroweak
  Precision Observables, New Physics and the Nature of a 126 GeV Higgs Boson}},
   {\em JHEP} {\bf 1308} (2013) 106,
  [\href{http://xxx.lanl.gov/abs/1306.4644}{{\tt arXiv:1306.4644}}].

\bibitem{Chen:2013kfa}
C.-Y. Chen, S.~Dawson, and C.~Zhang, {\it {Electroweak Effective Operators and
  Higgs Physics}},  {\em Phys.Rev.} {\bf D89} (2014), no.~1 015016,
  [\href{http://xxx.lanl.gov/abs/1311.3107}{{\tt arXiv:1311.3107}}].

\bibitem{Ciuchini:2014dea}
M.~Ciuchini, E.~Franco, S.~Mishima, M.~Pierini, L.~Reina, et~al., {\it {Update
  of the electroweak precision fit, interplay with Higgs-boson signal strengths
  and model-independent constraints on new physics}},
  \href{http://xxx.lanl.gov/abs/1410.6940}{{\tt arXiv:1410.6940}}.

\bibitem{Baak:2014ora}
{\bf Gfitter Group} Collaboration, M.~Baak et~al., {\it {The global electroweak
  fit at NNLO and prospects for the LHC and ILC}},  {\em Eur.Phys.J.} {\bf C74}
  (2014), no.~9 3046, [\href{http://xxx.lanl.gov/abs/1407.3792}{{\tt
  arXiv:1407.3792}}].

\bibitem{Durieux:2014xla}
G.~Durieux, F.~Maltoni, and C.~Zhang, {\it {Global approach to top-quark
  flavor-changing interactions}},  {\em Phys.Rev.} {\bf D91} (2015), no.~7
  074017, [\href{http://xxx.lanl.gov/abs/1412.7166}{{\tt arXiv:1412.7166}}].

\bibitem{Petrov:2015jea}
A.~A. Petrov, S.~Pokorski, J.~D. Wells, and Z.~Zhang, {\it {Role of low-energy
  observables in precision Higgs boson analyses}},  {\em Phys.Rev.} {\bf D91}
  (2015), no.~7 073001, [\href{http://xxx.lanl.gov/abs/1501.0280}{{\tt
  arXiv:1501.0280}}].

\bibitem{Wells:2014pga}
J.~D. Wells and Z.~Zhang, {\it {Precision Electroweak Analysis after the Higgs
  Boson Discovery}},  {\em Phys.Rev.} {\bf D90} (2014), no.~3 033006,
  [\href{http://xxx.lanl.gov/abs/1406.6070}{{\tt arXiv:1406.6070}}].

\bibitem{Ellis:2014dva}
J.~Ellis, V.~Sanz, and T.~You, {\it {Complete Higgs Sector Constraints on
  Dimension-6 Operators}},  {\em JHEP} {\bf 1407} (2014) 036,
  [\href{http://xxx.lanl.gov/abs/1404.3667}{{\tt arXiv:1404.3667}}].

\bibitem{Trott:2014dma}
M.~Trott, {\it {On the consistent use of Constructed Observables}},  {\em JHEP}
  {\bf 1502} (2015) 046, [\href{http://xxx.lanl.gov/abs/1409.7605}{{\tt
  arXiv:1409.7605}}].

\bibitem{Henning:2014wua}
B.~Henning, X.~Lu, and H.~Murayama, {\it {How to use the Standard Model
  effective field theory}},  \href{http://xxx.lanl.gov/abs/1412.1837}{{\tt
  arXiv:1412.1837}}.

\bibitem{Pomarol:2013zra}
A.~Pomarol and F.~Riva, {\it {Towards the Ultimate SM Fit to Close in on Higgs
  Physics}},  {\em JHEP} {\bf 1401} (2014) 151,
  [\href{http://xxx.lanl.gov/abs/1308.2803}{{\tt arXiv:1308.2803}}].

\bibitem{Falkowski:2014tna}
A.~Falkowski and F.~Riva, {\it {Model-independent precision constraints on
  dimension-6 operators}},  {\em JHEP} {\bf 1502} (2015) 039,
  [\href{http://xxx.lanl.gov/abs/1411.0669}{{\tt arXiv:1411.0669}}].

\bibitem{Grzadkowski:2010es}
B.~Grzadkowski, M.~Iskrzynski, M.~Misiak, and J.~Rosiek, {\it {Dimension-Six
  Terms in the Standard Model Lagrangian}},  {\em JHEP} {\bf 1010} (2010) 085,
  [\href{http://xxx.lanl.gov/abs/1008.4884}{{\tt arXiv:1008.4884}}].

\bibitem{Alonso:2013hga}
R.~Alonso, E.~E. Jenkins, A.~V. Manohar, and M.~Trott, {\it {Renormalization
  Group Evolution of the Standard Model Dimension Six Operators III: Gauge
  Coupling Dependence and Phenomenology}},  {\em JHEP} {\bf 1404} (2014) 159,
  [\href{http://xxx.lanl.gov/abs/1312.2014}{{\tt arXiv:1312.2014}}].

\bibitem{Buchmuller:1985jz}
W.~Buchmuller and D.~Wyler, {\it {Effective Lagrangian Analysis of New
  Interactions and Flavor Conservation}},  {\em Nucl.Phys.} {\bf B268} (1986)
  621.

\bibitem{Lehman:2014jma}
L.~Lehman, {\it {Extending the Standard Model Effective Field Theory with the
  Complete Set of Dimension-7 Operators}},  {\em Phys.Rev.} {\bf D90} (2014),
  no.~12 125023, [\href{http://xxx.lanl.gov/abs/1410.4193}{{\tt
  arXiv:1410.4193}}].

\bibitem{Z-Pole}
{The ALEPH, DELPHI, L3, OPAL, SLD Collaborations, the LEP Electroweak Working
  Group, the SLD Electroweak and Heavy Flavour Groups}, {\it {Precision
  Electroweak Measurements on the Z Resonance}},  {\em Phys. Rept.} {\bf 427}
  (2006) 257, [\href{http://xxx.lanl.gov/abs/hep-ex/0509008}{{\tt
  hep-ex/0509008}}].

\bibitem{Chivukula:1987py}
R.~S. Chivukula and H.~Georgi, {\it {Composite Technicolor Standard Model}},
  {\em Phys.Lett.} {\bf B188} (1987) 99.

\bibitem{Hall:1990ac}
L.~Hall and L.~Randall, {\it {Weak scale effective supersymmetry}},  {\em
  Phys.Rev.Lett.} {\bf 65} (1990) 2939--2942.

\bibitem{DAmbrosio:2002ex}
G.~D'Ambrosio, G.~Giudice, G.~Isidori, and A.~Strumia, {\it {Minimal flavor
  violation: An Effective field theory approach}},  {\em Nucl.Phys.} {\bf B645}
  (2002) 155--187, [\href{http://xxx.lanl.gov/abs/hep-ph/0207036}{{\tt
  hep-ph/0207036}}].

\bibitem{Buras:2000dm}
A.~Buras, P.~Gambino, M.~Gorbahn, S.~Jager, and L.~Silvestrini, {\it {Universal
  unitarity triangle and physics beyond the standard model}},  {\em Phys.Lett.}
  {\bf B500} (2001) 161--167,
  [\href{http://xxx.lanl.gov/abs/hep-ph/0007085}{{\tt hep-ph/0007085}}].

\bibitem{Manohar:1983md}
A.~Manohar and H.~Georgi, {\it {Chiral Quarks and the Nonrelativistic Quark
  Model}},  {\em Nucl.Phys.} {\bf B234} (1984) 189.

\bibitem{Jenkins:2013sda}
E.~E. Jenkins, A.~V. Manohar, and M.~Trott, {\it {Naive Dimensional Analysis
  Counting of Gauge Theory Amplitudes and Anomalous Dimensions}},  {\em
  Phys.Lett.} {\bf B726} (2013) 697--702,
  [\href{http://xxx.lanl.gov/abs/1309.0819}{{\tt arXiv:1309.0819}}].

\bibitem{Buchalla:2013eza}
G.~Buchalla, O.~Catá, and C.~Krause, {\it {On the Power Counting in Effective
  Field Theories}},  {\em Phys.Lett.} {\bf B731} (2014) 80--86,
  [\href{http://xxx.lanl.gov/abs/1312.5624}{{\tt arXiv:1312.5624}}].

\bibitem{Arzt:1994gp}
C.~Arzt, M.~Einhorn, and J.~Wudka, {\it {Patterns of deviation from the
  standard model}},  {\em Nucl.Phys.} {\bf B433} (1995) 41--66,
  [\href{http://xxx.lanl.gov/abs/hep-ph/9405214}{{\tt hep-ph/9405214}}].

\bibitem{Heinemeyer:2013tqa}
{\bf LHC Higgs Cross Section Working Group} Collaboration, S.~Heinemeyer
  et~al., {\it {Handbook of LHC Higgs Cross Sections: 3. Higgs Properties}},
  \href{http://xxx.lanl.gov/abs/1307.1347}{{\tt arXiv:1307.1347}}.

\bibitem{Buchalla:2014eca}
G.~Buchalla, O.~Cata, and C.~Krause, {\it {A Systematic Approach to the SILH
  Lagrangian}},  {\em Nucl.Phys.} {\bf B894} (2015) 602--620,
  [\href{http://xxx.lanl.gov/abs/1412.6356}{{\tt arXiv:1412.6356}}].

\bibitem{Jenkins:2013fya}
E.~E. Jenkins, A.~V. Manohar, and M.~Trott, {\it {On Gauge Invariance and
  Minimal Coupling}},  {\em JHEP} {\bf 1309} (2013) 063,
  [\href{http://xxx.lanl.gov/abs/1305.0017}{{\tt arXiv:1305.0017}}].

\bibitem{Wells:2005vk}
J.~D. Wells, {\it {TASI lecture notes: Introduction to precision electroweak
  analysis}},  \href{http://xxx.lanl.gov/abs/hep-ph/0512342}{{\tt
  hep-ph/0512342}}.

\bibitem{Agashe:2014kda}
{\bf Particle Data Group} Collaboration, K.~Olive et~al., {\it {Review of
  Particle Physics}},  {\em Chin.Phys.} {\bf C38} (2014) 090001.

\bibitem{Mohr:2012tt}
P.~J. Mohr, B.~N. Taylor, and D.~B. Newell, {\it {CODATA Recommended Values of
  the Fundamental Physical Constants: 2010}},  {\em Rev.Mod.Phys.} {\bf 84}
  (2012) 1527--1605, [\href{http://xxx.lanl.gov/abs/1203.5425}{{\tt
  arXiv:1203.5425}}].

\bibitem{Bardin:1999ak}
D.~Y. Bardin and G.~Passarino, {\it {The standard model in the making:
  Precision study of the electroweak interactions}}, .

\bibitem{Eichten:1983hw}
E.~Eichten, K.~D. Lane, and M.~E. Peskin, {\it {New Tests for Quark and Lepton
  Substructure}},  {\em Phys.Rev.Lett.} {\bf 50} (1983) 811--814.

\bibitem{Breit:1936zzb}
G.~Breit and E.~Wigner, {\it {Capture of Slow Neutrons}},  {\em Phys.Rev.} {\bf
  49} (1936) 519--531.

\bibitem{Altarelli:1989hv}
G.~Altarelli, R.~Kleiss, and C.~Verzegnassi, {\it {Z PHYSICS AT LEP-1.
  PROCEEDINGS, WORKSHOP, GENEVA, SWITZERLAND, SEPTEMBER 4-5, 1989. VOL. 1:
  STANDARD PHYSICS}}, .

\bibitem{Sirlin:1991fd}
A.~Sirlin, {\it {Theoretical considerations concerning the Z0 mass}},  {\em
  Phys.Rev.Lett.} {\bf 67} (1991) 2127--2130.

\bibitem{Isidori:2013cla}
G.~Isidori, A.~V. Manohar, and M.~Trott, {\it {Probing the nature of the
  Higgs-like Boson via $h \to V \mathcal{F}$ decays}},  {\em Phys.Lett.} {\bf
  B728} (2014) 131--135, [\href{http://xxx.lanl.gov/abs/1305.0663}{{\tt
  arXiv:1305.0663}}].

\bibitem{Isidori:2013cga}
G.~Isidori and M.~Trott, {\it {Higgs form factors in Associated Production}},
  {\em JHEP} {\bf 1402} (2014) 082,
  [\href{http://xxx.lanl.gov/abs/1307.4051}{{\tt arXiv:1307.4051}}].

\bibitem{Gonzalez-Alonso:2014eva}
M.~Gonzalez-Alonso, A.~Greljo, G.~Isidori, and D.~Marzocca, {\it
  {Pseudo-observables in Higgs decays}},  {\em Eur.Phys.J.} {\bf C75} (2015),
  no.~3 128, [\href{http://xxx.lanl.gov/abs/1412.6038}{{\tt arXiv:1412.6038}}].

\bibitem{L3entry}
O.~A. et~al. L3~Collab. {\em Phys. Rep. 236 1.} (1993).

\bibitem{Topaz}
T.~Collab. {\em Phys. Lett B34 7} (1995).

\bibitem{venus}
V.~Collab. {\em Phys. Lett B44 7} (1999).

\bibitem{Kirsch:1994cf1}
S.~Kirsch and T.~Riemann, {\it {SMATASY: A program for the model independent
  description of the Z resonance}},  {\em Comput.Phys.Commun.} {\bf 88} (1995)
  89--108, [\href{http://xxx.lanl.gov/abs/hep-ph/9408365}{{\tt
  hep-ph/9408365}}].

\bibitem{Kirsch:1994cf}
A.~Leike, T.~Riemann, and J.~Rose, {\it {S matrix approach to the Z line
  shape}},  {\em Phys.Lett.} {\bf B273} (1991) 513--518,
  [\href{http://xxx.lanl.gov/abs/hep-ph/9508390}{{\tt hep-ph/9508390}}].

\bibitem{Isidori:1993ms}
G.~Isidori, {\it {The Interference parameter in the model independent approach
  to Z line shape}},  {\em Phys.Lett.} {\bf B314} (1993) 139--148.

\bibitem{Freitas:2014hra}
A.~Freitas, {\it {Higher-order electroweak corrections to the partial widths
  and branching ratios of the Z boson}},  {\em JHEP} {\bf 1404} (2014) 070,
  [\href{http://xxx.lanl.gov/abs/1401.2447}{{\tt arXiv:1401.2447}}].

\bibitem{Group:2012gb}
{\bf CDF Collaboration, D0 Collaboration} Collaboration, T.~E.~W. Group, {\it
  {2012 Update of the Combination of CDF and D0 Results for the Mass of the W
  Boson}},  \href{http://xxx.lanl.gov/abs/1204.0042}{{\tt arXiv:1204.0042}}.

\bibitem{Awramik:2003rn}
M.~Awramik, M.~Czakon, A.~Freitas, and G.~Weiglein, {\it {Precise prediction
  for the W boson mass in the standard model}},  {\em Phys.Rev.} {\bf D69}
  (2004) 053006, [\href{http://xxx.lanl.gov/abs/hep-ph/0311148}{{\tt
  hep-ph/0311148}}].

\bibitem{Jenkins:2013zja}
E.~E. Jenkins, A.~V. Manohar, and M.~Trott, {\it {Renormalization Group
  Evolution of the Standard Model Dimension Six Operators I: Formalism and
  lambda Dependence}},  {\em JHEP} {\bf 1310} (2013) 087,
  [\href{http://xxx.lanl.gov/abs/1308.2627}{{\tt arXiv:1308.2627}}].

\bibitem{Jenkins:2013wua}
E.~E. Jenkins, A.~V. Manohar, and M.~Trott, {\it {Renormalization Group
  Evolution of the Standard Model Dimension Six Operators II: Yukawa
  Dependence}},  {\em JHEP} {\bf 1401} (2014) 035,
  [\href{http://xxx.lanl.gov/abs/1310.4838}{{\tt arXiv:1310.4838}}].

\bibitem{Alonso:2014zka}
R.~Alonso, H.-M. Chang, E.~E. Jenkins, A.~V. Manohar, and B.~Shotwell, {\it
  {Renormalization group evolution of dimension-six baryon number violating
  operators}},  {\em Phys.Lett.} {\bf B734} (2014) 302--307,
  [\href{http://xxx.lanl.gov/abs/1405.0486}{{\tt arXiv:1405.0486}}].

\bibitem{Brivio:2013pma}
I.~Brivio, T.~Corbett, O.~{\'E}boli, M.~Gavela, J.~Gonzalez-Fraile, et~al.,
  {\it {Disentangling a dynamical Higgs}},  {\em JHEP} {\bf 1403} (2014) 024,
  [\href{http://xxx.lanl.gov/abs/1311.1823}{{\tt arXiv:1311.1823}}].

\bibitem{GonzalezGarcia:1998ay}
M.~Gonzalez-Garcia, A.~Gusso, and S.~Novaes, {\it {Constraints on four fermion
  contact interactions from precise electroweak measurements}},  {\em J.Phys.}
  {\bf G24} (1998) 2213--2221,
  [\href{http://xxx.lanl.gov/abs/hep-ph/9802254}{{\tt hep-ph/9802254}}].

\end{thebibliography}\endgroup

\end{document}